\documentclass[12pt]{article}
\pdfoutput=1
\usepackage{color}
\definecolor{darkblue}{rgb}{0.1,0.1,.7}
\usepackage[colorlinks, linkcolor=darkblue, citecolor=darkblue, urlcolor=darkblue, linktocpage]{hyperref}
\usepackage[]{amsmath}
\usepackage[]{graphicx}
\usepackage[]{latexsym}
\usepackage{slashed,graphicx,color,amsmath,amssymb}
\usepackage[mathscr]{eucal}
\usepackage{mathrsfs}
\usepackage{geometry}
\usepackage[margin=10pt,font=small,labelfont=bf]{caption}
\usepackage{amscd}
\usepackage{bm}
\usepackage{xcolor}
\usepackage{upgreek}
\usepackage[square, comma, sort&compress,numbers]{natbib}
\usepackage[all,cmtip]{xy}
\numberwithin{equation}{section}
\newcommand{\rB}{\mathrm{B}}
\newcommand{\rC}{\mathrm{C}}
\newcommand{\rD}{\mathrm{D}}

\newcommand{\rK}{\mathrm{K}}

\newcommand{\rT}{\mathrm{T}}

\begin{document}
\vspace*{-.6in} \thispagestyle{empty}
%\begin{flushright}
%\textcolor{gray}{Personal Notes}
%\end{flushright}
\vspace{.2in} {\Large
\begin{center}
{\bf Entanglement Entropy in Integrable Field Theories\\ with Line Defects II. Non-topological Defect}
\end{center}}
\vspace{.2in}
\begin{center}
Yunfeng Jiang
\\
\vspace{.3in}
\small{\textit{Institut f{\"u}r Theoretische Physik,
ETH Z{\"u}rich}},\\
\small{\textit{
Wolfgang Pauli Strasse 27,
CH-8093 Z{\"u}rich, Switzerland}
}

\end{center}

\vspace{.3in}

\begin{abstract}
\normalsize{This is the second part of two papers where we study the effect of integrable line defects on bipartite entanglement entropy in integrable field theories. In this paper, we consider non-topological line defects in Ising field theory. We derive an infinite series expression for the entanglement entropy and show that both the UV and IR limits of the bulk entanglement entropy are modified by the line defect. In the UV limit, we give an infinite series expression for the coefficient in front of the logarithmic divergence and the exact defect $g$-function. By tuning the defect to be purely transmissive and reflective, we recover correctly the entanglement entropy of the bulk and with integrable boundary respectively.}
\end{abstract}

\vskip 1cm \hspace{0.7cm}

\newpage

\setcounter{page}{1}
\begingroup
\hypersetup{linkcolor=black}
\tableofcontents
\endgroup

%%%%%%%%%%%%%%%%%%%%%%%%%%%%%%%%%%%%%%%%%%%%%%%%%%%%%%%%%%%%%%%%%%%%%%%%
\section{Introduction}
\label{sec:intro}
%%%%%%%%%%%%%%%%%%%%%%%%%%%%%%%%%%%%%%%%%%%%%%%%%%%%%%%%%%%%%%%%%%%%%%%%
We continue our study of the effects of integrable line defects on bipartite entanglement entropy (EE) in integrable field theories initiated in \cite{Jiang:2017qhn}. We have studied the effects of topological defects on EE in interacting field theories in \cite{Jiang:2017qhn}. In the current work, we focus on the effects of \emph{non-topological defects}. Due to the no-go theorem \cite{Delfino:1994nr}, integrable non-topological defects are only allowed in free theories with the $S$-matrix being $S=\pm1$. Here we consider the free fermion case and take the \emph{massive} Ising field theory with a non-topological defect as our main example.

The entanglement entropy of free fermion with defects have been investigated extensively in different frameworks. In \cite{Brehm:2015lja} (see also \cite{Sakai:2008tt} for the bosonic case), the authors computed the entanglement entropy across a conformal interface using the folding trick and boundary CFT techniques. Their results are in agreement with a series of works on EE in critical chains with defects \cite{Igloi,Eisler2010,Eisler2012,Eisler2012Evol}. In all these cases, the EE takes the following form
\begin{align}
\label{eq:previousEE}
S=\sigma(\kappa)\,\log L+C
\end{align}
where $\kappa$ is a parameter characterizing the defect and $L$ is the system size on one side of the defect. Here $\sigma(\kappa)$ is a non-trivial function of $\kappa$ and $C$ is a constant that does not depend on $\kappa$ or $L$. A very similar formula has been found for spatial EE of 1D free gas and quantum wire junctions \cite{Calabrese:2011zzb,Calabrese:2011ru} where the system size $L$ in (\ref{eq:previousEE}) is replaced by the number of particles $N$.

\begin{figure}[h!]
\begin{center}
\includegraphics[scale=0.3]{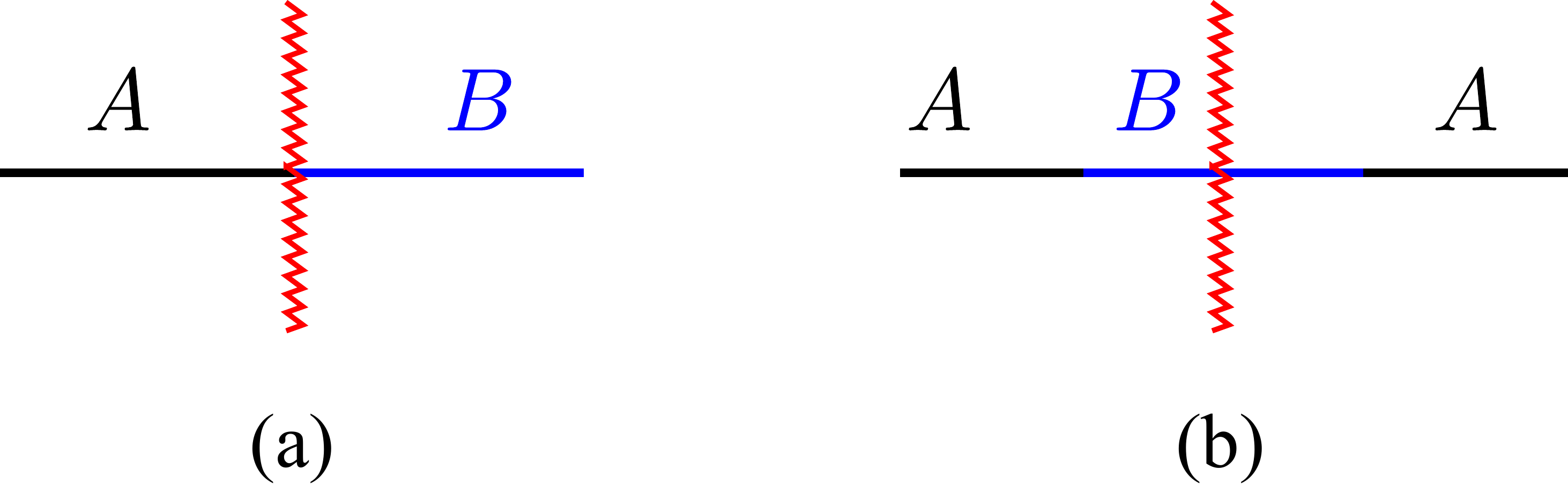}
\caption{Different set-ups for entanglement entropy. Figure (a) is the set-up of previous studies where the subsystem size goes to infinity and the defect is localized at one end of the subsystem. Figure (b) is our set-up where the subsystem is finite with length $r$ and the defect is located within the interval.}
\label{fig:diff_set-up}
\end{center}
\end{figure}

Before embarking on explicit calculations, let us point out the crucial differences between the current paper and the previous works mentioned above. First and foremost, all the previous studies focus on the critical point while we consider the \emph{massive} Ising field theory which describes the free fermion away from criticality. Moreover, the set-ups of the computation of EE are different. In the previous cases, the subsystem is taken as the whole system on one side of the defect and the defect is located on one end of the subsystem. On the other hand, we consider a finite interval with length $r$ as the subsystem and place the defect within the interval, as is shown in figure\,\ref{fig:diff_set-up}.
Finally our method is different. It is based heavily on the integrability of the model. The bulk physics and the defect are characterized by the bulk scattering amplitudes and the scattering amplitudes of bulk particles on the defect, respectively. Due to these differences, it is expected that our result, which we will present shortly, is different from the previously studied cases.

The configuration we consider is given by figure\,\ref{fig:diff_set-up} (b). For topological defects, the result does not depend on the exact position of the defect within the interval. On the contrary, for non-topological defects the result depends on the position of the defect. Let us denote the length of the interval by $r$ and the distance between the defect and left (right) end of the interval by $r_L\, (r_R)$. We have $r_L+r_R=r$. Denoting the UV cut-off by $\varepsilon$, our result can be summarized as the following. For $\varepsilon\ll r_L,r_R\ll \mathrm{m}^{-1}$ where $\mathrm{m}$ is the mass scale of the theory, the entanglement entropy can be written in the following form\footnote{This form is derived directly from the UV limit of the spectral expansion and it highlights the property of non-topological defect which is transmissive and reflective simultaneously. It can also be written in a more compact form, see (\ref{eq:simpleEE}) where the effect of the defect on EE is contained in the constant term.}
\begin{align}
\label{eq:SUV}
S_A^{\text{defect}}=&\,f(\chi)\log(r/\varepsilon)+f_L(\chi)\log(2r_L/\varepsilon)+f_R(\chi)\log(2r_R/\varepsilon)\\\nonumber
&\,+g(\chi)+g_L(\chi)+g_R(\chi)
\end{align}
where $\chi$ is a parameter that characterizes the defect and is defined in (\ref{eq:defectLagrange}) and (\ref{eq:chi-g}) in the Lagrangian formulation. For $r_L,r_R\gg \mathrm{m}$, we have
\begin{align}
\label{eq:SIR}
S_A^{\text{defect}}=-\frac{c}{3}\log(\varepsilon\mathrm{m})+U+\mathcal{O}(e^{-a\,\mathrm{m}r})
\end{align}
where the universal constant $U$ is already defined and computed in the bulk case for the Ising model in \cite{Cardy:2007mb} with the value $U=-0.131984..$\footnote{In the IR limit, the defect only modifies the exponential corrections of EE and has no effect on the constant piece, so we take the same universal constant $U$ as in the bulk case.}. In the IR limit, the non-topological defect only affects the exponential corrections of EE and have no effect on the constant term, which is similar to the topological defect. The leading exponential corrections will be computed in section\,\ref{sec:IRlimit}. The defect under consideration is parity invariant and we have $f_L(\chi)=f_R(\chi)$ and $g_L(\chi)=g_R(\chi)$. There are two types of logarithmic divergences in (\ref{eq:SUV}). The first one depends on $r$ which is similar to the case of the bulk \cite{Cardy:2007mb} and the case with a topological defect \cite{Jiang:2017qhn}. The second type depends on $2r_L$ or $2r_R$ and takes the same form as the logarithmic divergence in the integrable boundary case \cite{CastroAlvaredo:2008pf}. The coefficients in front of the logarithmic divergences depend on the parameter $\chi$ in a non-trivial way. In the limit $\chi=0$ the defect becomes transparent, we have $f_L(0)=f_R(0)=0$, $f(0)=c/3$ and we recover the bulk entanglement entropy. On the other hand, for $\chi=\pi/2$, the defect becomes non-transmissive which means no particle can go through the line defect. We have in this case $f(\pi/2)=0$ and $f_L(\pi/2)=f_R(\pi/2)=c/6$ and the result is simply a sum of entanglement entropies of two integrable boundary systems, one on each side of the defect. For generic $\chi$, all the three coefficients are non-vanishing but the relation $f(\chi)+f_L(\chi)+f_R(\chi)=c/3$ holds. This will be proved in section\,\ref{sec:EEdefect}. We can view the result of EE in the presence of non-topological defect as an interpolation between those of the bulk case and the integrable boundary case.\par

In the second line of (\ref{eq:SUV}), there are three functions depending on the parameter $\chi$. These functions are essentially the boundary entropy or $g$-function which measure the boundary and defect degrees of freedom \cite{Affleck:1991tk}. The functions $g_L(\chi)$ and $g_R(\chi)$ are the boundary $g$-functions and $g(\chi)$ is the defect $g$-function.\par

Using the relation $f(\chi)+f_L(\chi)+f_R(\chi)=c/3$, we can rewrite (\ref{eq:SUV}) in a simpler form
\begin{align}
\label{eq:simpleEE}
S_A^{\text{defect}}=\frac{c}{3}\log\left(r/\varepsilon\right)+G(\chi)
\end{align}
where
\begin{align}
G(\chi)=f_L(\chi)\log(2r_L/r)+f_R(\chi)\log(2r_R/r)+g(\chi)+g_L(\chi)+g_R(\chi).
\end{align}
Comparing our result (\ref{eq:simpleEE}) with the previously studied cases (\ref{eq:previousEE}), we find that the coefficient of the logarithmic divergent term is the same constant as in the bulk case. This is mainly due to our set-up where we place the defect within the interval while in the other cases the defect is placed at one end of the subsystem. In both cases, the defects lead to a contribution of order $\mathcal{O}(1)$ in EE. The difference is that in the previously studied cases, the $\mathcal{O}(1)$ term is a constant independent of the parameter $\kappa$ that characterize the defect and the size of the subsystem $L$ while in our case the $\mathcal{O}(1)$ term $G(\chi)$ depends on both $\chi$ and $r,r_L,r_R$.

We can have another situation in the non-topological defect case where both UV and IR limits are present. More precisely, we can take $\varepsilon\ll r_L\ll \mathrm{m}^{-1}$, $r_R\gg \mathrm{m}^{-1}$ or $\varepsilon\ll r_R\ll \mathrm{m}^{-1}$, $r_L\gg \mathrm{m}^{-1}$. For $\varepsilon\ll r_L\ll\mathrm{m}^{-1}, r_R\gg\mathrm{m}^{-1}$ we have
\begin{align}
S_A^{\text{defect}}=&\,f_L(\chi)\log(2r_L/\varepsilon)+g_L(\chi)-\left(\frac{c}{3}-f_L(\chi)\right)\log(\varepsilon \mathrm{m})+U+\cdots\\\nonumber
=&\,-\frac{c}{3}\log(\varepsilon\mathrm{m})+f_L(\chi)\log(2mr_L)+g_L(\chi)+U+\cdots
\end{align}
where the ellipsis denote the terms that are decaying exponentially in $mr_R$ and with a power law in $mr_L$. Exchanging ``$L$'' and ``$R$'' gives the result of the other case. As we can see explicitly, the behavior of both UV and IR limits are present in this case.

The rest of the paper is structured as follows. In section \ref{sec:defectIsing}, we review some basic facts about the defect Ising model and the corresponding replica theory. In section \ref{sec:EEdefect}, we apply the replica trick and the form factor approach to compute the defect EE. Our main result is an infinite series expression for the defect EE. Using this infinite series expression, we consider various asymptotic behaviors of the entanglement entropy in section \ref{sec:asymp}, obtaining infinite series expressions for $f(\chi)$, $f_L(\chi)$ and $f_R(\chi)$. In section \ref{sec:g-function}, we derive infinite series expressions for the boundary/defect $g$-functions $g(\chi)$, $g_L(\chi)$ and $g_R(\chi)$. We conclude in section \ref{sec:conclusion}.

%%%%%%%%%%%%%%%%%%%%%%%%%%%%%%%%%%%%%%%%%%%%%%%%%%%%%%%%%%%%%%%%%%%%%%%%
\section{Defect Ising model}
\label{sec:defectIsing}
%%%%%%%%%%%%%%%%%%%%%%%%%%%%%%%%%%%%%%%%%%%%%%%%%%%%%%%%%%%%%%%%%%%%%%%%
In this section, we briefly review some basic facts about the defect Ising field theory following mainly \cite{Delfino:1994nr}. The bulk theory is the massive Ising field theory whose excitations are given by massive Majorana fermions with the $S$-matrix $S=-1$. The bulk Lagrangian density is
\begin{align}
\mathcal{L}_\rB=\overline{\Psi}(x,t)(i\gamma^\mu\partial_\mu-\mathrm{m})\Psi(x,t)
\end{align}
where
\begin{align}
\gamma^0=\sigma_2,\qquad \gamma^1=-i\sigma_1
\end{align}
and the fermionic field $\Psi(x,t)$ is real, namely $\Psi^\dagger(x,t)=\Psi(x,t)$. The line defect of Ising model can be introduced by the following defect term
\begin{align}
\label{eq:defectLagrange}
\mathcal{L}_\rD=-g\,\delta(x)\overline{\Psi}(x,t)\Psi(x,t).
\end{align}
where $g$ is a coupling constant. This defect can be transmissive and reflective at the same time. It is useful to describe the interaction between the defect and bulk particles by an algebra called the defect algebra, which is akin to the Zamolodchikov-Faddeev algebra. The defect algebra \cite{Jiang:2017qhn} for defect Ising model reads
\begin{align}
A^\dagger(\theta)D=&\,R(\theta)A^\dagger(-\theta)D+T(\theta)D A^\dagger(\theta),\\\nonumber
DA^\dagger(\theta)=&\,R(-\theta)DA^\dagger(-\theta)+T(-\theta)A^\dagger(\theta)D.
\end{align}
The consistency of the defect algebra leads to a set of constraints for the transmission and reflection amplitudes. The transmission and reflection amplitudes of the defect Ising field theory have been worked out in \cite{Delfino:1994nr}
\begin{align}
T(\theta,\chi)=\frac{\cos\chi\sinh\theta}{\sinh\theta-i\sin\chi},\qquad R(\theta,\chi)=i\frac{\sin\chi\cosh\theta}{\sinh\theta-i\sin\chi}
\end{align}
where the parameter $\chi$ is related to the coupling constant $g$ in (\ref{eq:defectLagrange}) via
\begin{align}
\label{eq:chi-g}
\sin\chi=-\frac{4g}{g^2+4}.
\end{align}
It is easy to see that when $g=\pm2$, $T(\theta,\chi)=0$, $R(\theta,\chi)=1$ and the defect becomes purely reflective.\par

In order to compute the entanglement entropy, we apply the replica trick \cite{Cardy:2007mb}. In this approach, one introduces $n$ copies of the theory, glued together cyclicly through the cuts. In the replica theory, one can define new operators called the branch-point twist fields and the computation of entanglement entropy essentially reduces to the computation of the two-point function of branch-point twist fields. In the presence of integrable defects or boundaries, these extra structure are also duplicated. The calculation of entanglement entropy in these cases are reduced to the computation of two-point function of branch-point twist fields in the presence of defects or boundaries. For more details, we refer to \cite{Jiang:2017qhn}.

Computing such correlation functions in massive field theories is highly non-trivial. When the theory is integrable, we can apply the powerful method of form factor bootstrap \cite{Smirnov:1992vz,Karowski:1978vz}. In order to compute the defect two-point function, we need to know the form factors of the local operators and the matrix elements of the defect operator \cite{Jiang:2017qhn}. The form factors of branch-point twist fields of Ising field theory have been determined in \cite{Cardy:2007mb,CastroAlvaredo:2008pf}. Form factors of particles with different replica numbers are related to the one with the same replica number by
\begin{align}
\label{eq:replicaN}
F_k^{\mathcal{T}|a_1\cdots a_k}(\theta_1,\cdots,\theta_k)=F_k^{\mathcal{T}|1\cdots 1}(\theta_1+2\pi i(a_1-1),\cdots,\theta_k+2\pi i(a_k-1))
\end{align}
where $1\le a_i\le n$ are the replica numbers. The multi-particle form factors $F_k^{\mathcal{T}|1\cdots1}(\theta_1,\cdots,\theta_k)$ can be written in terms of two-particle form factors as a Pfaffian,
\begin{align}
\label{eq:Pfaffian}
F_k^{\mathcal{T}|1\cdots1}(\theta_1,\cdots,\theta_k)=\langle\mathcal{T}\rangle\,\text{Pf}(\hat{K})
\end{align}
where
\begin{align}
\hat{K}_{ij}=K(\theta_i-\theta_j)=\frac{F_{\text{min}}^{\mathcal{T}|11}(\theta_i-\theta_j)}{F_{\text{min}}^{\mathcal{T}|11}(i\pi)P(\theta_i-\theta_j)}
\end{align}
and
\begin{align}
F_{\text{min}}^{\mathcal{T}|11}(\theta)=-i\sinh\left(\frac{\theta}{2n}\right),\qquad P(\theta)=\frac{2n\sinh\left( \frac{i\pi+\theta}{2n} \right)\sinh\left( \frac{i\pi-\theta}{2n} \right)}{\sinh\left( \frac{\pi}{n} \right)}.
\end{align}
The function $K(\theta)$ satisfies the following properties
\begin{align}
\label{eq:propertyK}
K(\theta)=-K(-\theta),\qquad \left.K(\theta)\right|_{n=1}=0,\qquad K(\theta+is)^*=-K(\theta-is).
\end{align}
Using these properties, it is straightforward to show that
\begin{align}
\label{eq:Fconj}
\left(F_k^{\mathcal{T}|a_1\cdots a_k}(\theta_1,\cdots,\theta_k)\right)^*=F_k^{\mathcal{T}|a_1\cdots a_k}(-\theta_1,\cdots,-\theta_k).
\end{align}

\subsection{Defect matrix elements}
Another important ingredient for our computation is the matrix elements of the defect operator. This operator can be written in terms of Zamolodchikov-Faddeev operators \cite{CastroAlvaredo:2002dj}, similar to the integrable boundary state \cite{Ghoshal:1993tm}. The matrix elements of the defect operator can be determined following the method in \cite{Delfino:1994nr}. We refer to appendix \ref{sec:defectO} for more details and an explicit form of the defect operator in the replica theory. All the matrix elements can be written in terms of the three fundamental building blocks
\begin{align}
\label{eq:D12}
\rD_{1,1}^{\mu,\nu}=&\,{_\mu\langle}\theta|\mathbb{D}|\theta'\rangle_\nu=2\pi\,\hat{T}(\theta){\delta_{\mu,\nu}}\,\delta(\theta-\theta'),\\\nonumber
\rD_{2,0}^{\mu_1\mu_2,}=&\,{_{\mu_1\mu_2}\langle}\theta_1,\theta_2|\mathbb{D}|0\rangle=2\pi\,\hat{R}(\theta_1){\delta_{\mu_1,\mu_2}}\,\delta(\theta_1+\theta_2),\\\nonumber
\rD_{0,2}^{,\nu_1\nu_2}=&\,\langle0|\mathbb{D}|\theta'_1,\theta'_2\rangle_{\nu_1\nu_2}=2\pi\,\hat{R}(\theta'_1){\delta_{\nu_1,\nu_2}}\,\delta(\theta'_1+\theta'_2).
\end{align}
where
\begin{align}
\label{eq:RThat}
\hat{T}(\theta)=&\,T\left(\frac{i\pi}{2}-\theta,\chi\right)=\frac{\cos\chi\cosh\theta}{\cosh\theta-\sin\chi},\\\nonumber
\hat{R}(\theta)=&\,R\left(\frac{i\pi}{2}-\theta,\chi\right)=-\frac{i\sin\chi\sinh\theta}{\cosh\theta-\sin\chi}
\end{align}
and $\mu_i,\nu_i=1,\cdots,n$ are replica numbers. Note that only the particles with the same replica number can interact with each other\footnote{By this we simply mean exchanging the order of two particles with the same number gives a minus sign.}. The quantities $\hat{T}(\theta)$ and $\hat{R}(\theta)$ are the Wick rotated transmission and reflection amplitudes. They satisfy the following properties
\begin{align}
\label{eq:parity}
\hat{T}(-\theta)=\hat{T}(\theta),\qquad \hat{R}(-\theta)=-\hat{R}(\theta).
\end{align}
In what follows, we will omit the replica numbers on the states to avoid clutter, but it should be understood that each excitation is labeled by a replica number. The general defect matrix element can be obtained by recursion relations. For fermionic case, the recursion relations are
\begin{align}
\label{eq:rec1}
&\,\langle\theta_1,\cdots,\theta_m,\textcolor{red}{\theta}|\mathbb{D}|\theta'_1,\cdots,\theta'_n\rangle\\\nonumber
&\,=2\pi\sum_{i=1}^m(-1)^{\delta_{\mu_i,\mu_{i+1}}+\cdots+\delta_{\mu_i,\mu_m}}
\hat{R}(\textcolor{red}{-\theta}){\delta_{\mu,\mu_i}}\delta(\textcolor{red}{\theta}+\theta_i)
\langle\theta_1,\cdots,\theta_{i-1},\theta_{i+1},\cdots,\theta_m|\mathbb{D}|\theta'_1,\cdots,\theta'_n\rangle\\\nonumber
&\,+2\pi\sum_{j=1}^n(-1)^{\delta_{\nu_1,\nu_j}+\cdots+\delta_{\nu_{j-1},\nu_j}}\hat{T}(\textcolor{red}{\theta}){\delta_{\mu,\nu_j}}\delta(\textcolor{red}{\theta}-\theta'_j)
\langle\theta_1,\cdots,\theta_m|\mathbb{D}|\theta'_1,\cdots,\theta'_{j-1},\theta'_{j+1},\cdots,\theta_n\rangle.
\end{align}
and
\begin{align}
\label{eq:rec2}
&\,\langle\theta_1,\cdots,\theta_m|\mathbb{D}|\textcolor{red}{\theta},\theta'_1,\cdots,\theta'_n\rangle\\\nonumber
&\,=2\pi\sum_{i=1}^n(-1)^{\delta_{\nu_1,\nu_i}+\cdots\delta_{\nu_{i-1},\nu_i}}\hat{R}(\textcolor{red}{\theta}){\delta_{\mu,\nu_i}}\delta(\textcolor{red}{\theta}+\theta'_i)
\langle\theta_1,\cdots,\theta_m|\mathbb{D}|\theta'_1,\cdots,\theta'_{i-1},\theta'_{i+1},\cdots,\theta'_n\rangle\\\nonumber
&\,+2\pi\sum_{j=1}^m(-1)^{\delta_{\mu_j,\mu_{j+1}}+\cdots+\delta_{\mu_j,\mu_m}}\hat{T}(\textcolor{red}{\theta}){\delta_{\mu,\mu_j}}\delta(\textcolor{red}{\theta}-\theta_j)
\langle\theta_1,\cdots,\theta_{j-1},\theta_{j+1},\cdots,\theta_m|\mathbb{D}|\theta'_1,\cdots,\theta'_n\rangle.
\end{align}
where as a convention we define $\delta_{\mu_0,\mu_j}=\delta_{\nu_0,\nu_j}=0$. Removing all the factors $(-1)^m$ from (\ref{eq:rec1}) and (\ref{eq:rec2}) gives the bosonic recursion relations.\par

%%%%%%%%%%%%%%%%%%%%%%%%%%%%%%%%%%%%%%%%%%%%%%%%%%%%%%%%%%%%%%%%%%%%%%%%
\section{Entanglement entropy in the presence a non-topolotical defect}
\label{sec:EEdefect}
%%%%%%%%%%%%%%%%%%%%%%%%%%%%%%%%%%%%%%%%%%%%%%%%%%%%%%%%%%%%%%%%%%%%%%%%
Using the replica trick, the bipartite entanglement entropy is given by $\langle\mathcal{T}(t_1)\mathbb{D}(t_\rD)\tilde{\mathcal{T}}(t_2)\rangle$ with $t_1<t_\rD<t_2$. We have $|t_1-t_\rD|=r_L$, $|t_2-t_\rD|=r_R$ and $|t_1-t_2|=r$. The quantity looks like a three-point function with the third operator being an extended object --- the line defect operator. In the UV limit where $r,r_L,r_R\to 0$, the defect two-point function behaves as
\begin{align}
\langle\mathcal{T}(t_1)\mathbb{D}(t_\rD)\tilde{\mathcal{T}}(t_2)\rangle=\frac{\mathcal{N}}{r_L^{\Delta_L}\,r_R^{\Delta_R}\,r^\Delta}
\end{align}
where the dimensions $\Delta(\chi),\Delta_L(\chi),\Delta_R(\chi)$ depend on $\chi$
and $\mathcal{N}$ is a normalization factor. We choose the normalization for the defect operator such that $\langle\mathbb{D}\rangle=1$. Since the defect operator is dimensionless\footnote{This is obvious from its explicit form in terms of ZF operators given in (\ref{eq:Dd1})}, by dimensional analysis we have
\begin{align}
\label{eq:delta}
\Delta(\chi)+\Delta_L(\chi)+\Delta_R(\chi)=4\Delta_n,\qquad \Delta_n=\frac{c}{24}\left(n-\frac{1}{n}\right).
\end{align}
It follows immediately that
\begin{align}
\label{eq:fff}
f(\chi)+f_L(\chi)+f_R(\chi)=\frac{c}{3}.
\end{align}
We will find that this relation is consistent with the explicit calculation by the form factor approach in section\,\ref{sec:asymUV}. The scaling dimensions $\Delta(\chi)$, $\Delta_L(\chi)=\Delta_R(\chi)$ can be computed by the form factor method \cite{Babujian:2003sc,Bianchini:2015uea}, but the results are given in terms of infinite sums inherited from the spectral expansion. It is thus very desirable to have a closed expression for these functions by defect CFT calculations. In terms of the defect two-point function, the entanglement entropy is given by
\begin{align}
S_A^{\text{defect}}=&\,-\lim_{n\to1}\frac{d}{dn}\mathcal{Z}_n\varepsilon^{\Delta+\Delta_L+\Delta_R}
\langle\mathcal{T}(t_1)\mathbb{D}(t_\rD)\tilde{\mathcal{T}}(t_2)\rangle\\\nonumber
=&\,-\lim_{n\to1}\frac{d}{dn}\mathcal{Z}_n\varepsilon^{4\Delta_n}
\langle\mathcal{T}(t_1)\mathbb{D}(t_\rD)\tilde{\mathcal{T}}(t_2)\rangle
\end{align}

\subsection{Spectral expansion}
The spectral expansion of the defect two-point function is given by
\begin{align}
\langle\mathcal{T}(t_1)\mathbb{D}(t_\rD)\tilde{\mathcal{T}}(t_2)\rangle=\sum_{M,N=0}^\infty f_{M,N}
\end{align}
where
\begin{align}
\label{eq:fMN}
f_{M,N}=&\,\frac{1}{M!N!}\sum_{j_1,\cdots,j_M=1}^n\sum_{k_1,\cdots,k_N=1}^n\int_{-\infty}^\infty\prod_{r=1}^M\frac{d\theta_r}{2\pi}
\prod_{s=1}^N\frac{d\theta'_s}{2\pi}\rD_{M,N}\\\nonumber
&\,\times F_M^{\mathcal{T}|j_1\cdots j_M}(\theta_1,\cdots,\theta_M)\left(F_N^{\mathcal{T}|k_1\cdots k_N}(\theta'_1,\cdots,\theta'_N)\right)^*
\times e^{-(\mathrm{m} r_L\sum_{r=1}^M\cosh\theta_r+\mathrm{m} r_R\sum_{s=1}^N\cosh\theta'_s)}.
\end{align}
Here $n$ is the number of replica and $\rD_{M,N}$ are the defect matrix elements
\begin{align}
\rD_{M,N}=\langle\theta_M,\cdots,\theta_1|\mathbb{D}|\theta'_1,\cdots,\theta'_N\rangle.
\end{align}
For an integrable defect, the defect matrix elements $\rD_{M,N}$ are non-zero for $M+N=0$ (mod 2). we have $f_{1,N}=f_{M,1}=0$ due to the fact that $F_1^{\mathcal{T}}=0$.\par

The matrix elements $\rD_{M,N}$ can be evaluated explicitly using the recursion relations (\ref{eq:rec1}) and (\ref{eq:rec2}). The number of terms grows quickly with the number of particles $M$ and $N$. The defect matrix elements are given in terms of transmission and reflection amplitudes $\hat{T}$ and $\hat{R}$ and $\delta$-functions. After integrating out the $\delta$-functions, we obtain a simpler multiple integral representation in the spectral expansion. In what follows, we introduce a diagrammatic representation for each term in the spectral expansion which makes it easier to write down the integrand (\ref{eq:fMN}).\par

\begin{figure}
\begin{center}
\includegraphics[scale=0.7]{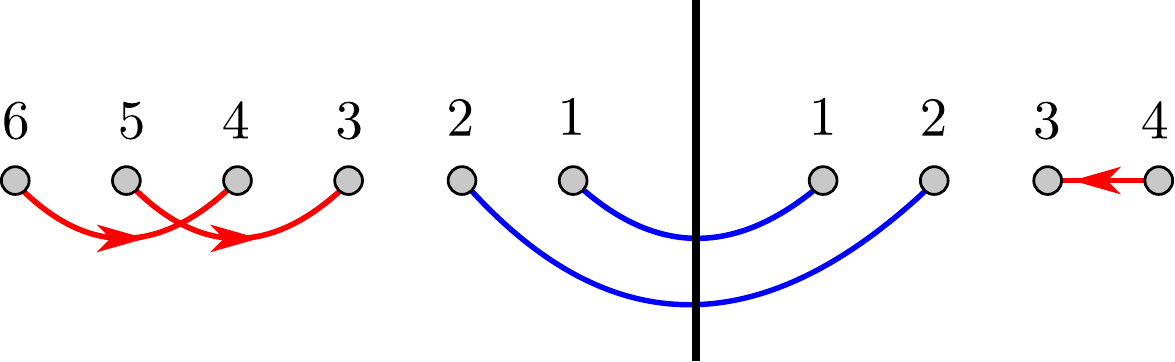}
\caption{An example of diagrammatic representation. This term is one of the terms of $f_{6,4}$.}
\label{fig:DT}
\end{center}
\end{figure}

\begin{figure}[h!]
\begin{center}
\includegraphics[scale=0.7]{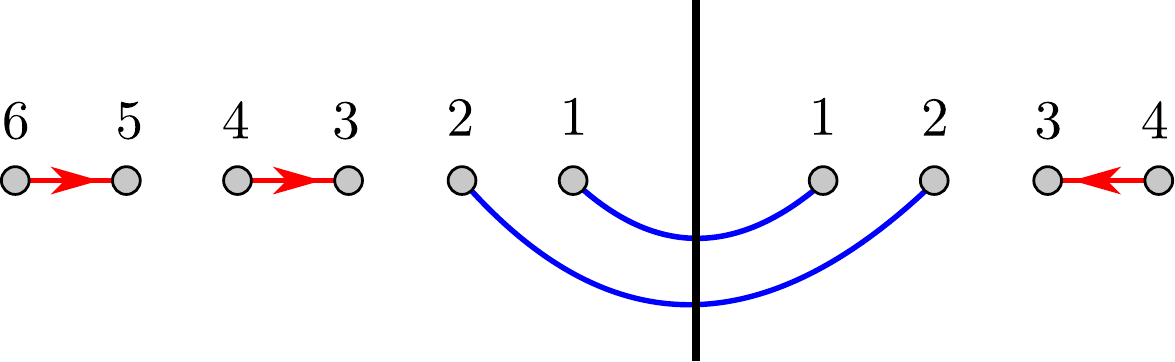}
\caption{The standard diagram $\rT_{6,4}^{(2)}$.}
\label{fig:dipole}
\end{center}
\end{figure}

Let us denote the particles on the left- and right- hand side of the defect by $\{\theta_a,j_a\}$ and $\{\theta'_b,j'_b\}$ respectively. Here $\theta_a$, $\theta'_b$ are rapidities of the particles and $j_a,j'_b$ are the replica numbers of the particles. For the integrand of $f_{M,N}$, $a=1,\cdots,M$ and $b=1,\cdots,N$. In the presence of an integrable defect, the particles have to be paired. This follows from the constraints imposed by the $\delta$-functions in the defect matrix element. A pair of two particles on the same side of the defect leads to a reflection amplitude while a pair of two particles on different sides of the defect leads to a transmission amplitude. In order to explain the rules, let us consider an example given in figure\,\ref{fig:DT}.
The pairs of the particles on the same sides are linked by directed red lines. Each directed link $\{\theta_a,j_a\}\to\{\theta_b,j_b\}$ corresponds to the factor
\begin{align}
\{\theta_a,j_a\}\to\{\theta_b,j_b\}:&\,\qquad \hat{R}(-\theta_b)\delta(\theta_a+\theta_b)\delta_{j_a,j_b}=\hat{R}(\theta_a)\delta(\theta_a+\theta_b)\delta_{j_a,j_b},\\\nonumber
\{\theta'_a,j'_a\}\to\{\theta'_b,j'_b\}:&\,\qquad \hat{R}(\theta_b)\delta(\theta_a+\theta_b)\delta_{j_a,j_b}
\end{align}
The direction of the arrows are always pointing to the defect line. The pair that involves particles on both sides are linked by a blue line and corresponds to the factor
\begin{align}
\{\theta_a,j_b\}\leftrightarrow\{\theta'_b,j'_b\}:\qquad \hat{T}(\theta_a)\delta(\theta_a-\theta'_b)\delta_{j_a,j'_b}.
\end{align}
When two lines cross, it corresponds to the $S$-matrix, which is simply $(-1)^{\delta_{j_aj_b}}$ for the replica theory. These rules give us the main part of the integrand. After writing down all the factors, we need to sum over all the replica indices and integrate over all rapidities. Following these rules and taking into account the exponential factors, the term corresponds to figure\,\ref{fig:DT} is given by
\begin{align}
\texttt{term}=&\,\frac{1}{6!\,4!}\sum_{j_1,\cdots,j_6=1}^n\sum_{j'_1,\cdots,j'_4=1}^n
\int\prod_{r=1}^6\frac{d\theta_r}{2\pi}\prod_{s=1}^4\frac{d\theta'_s}{2\pi}\,
\hat{R}(\theta_5)\hat{R}(\theta_6)\hat{R}(\theta'_3)\hat{T}(\theta_1)\hat{T}(\theta_2)(-1)^{\delta_{j_5,j_6}}\\\nonumber
&\times\delta(\theta_3+\theta_5)\delta(\theta_4+\theta_6)\delta(\theta'_3+\theta'_4)\delta(\theta_1-\theta'_1)\delta(\theta_2-\theta'_2)
\times\delta_{j_3,j_5}\delta_{j_4,j_6}\delta_{j'_3,j'_4}\delta_{j_1,j'_1}\delta_{j_2,j'_2}\\\nonumber
&\times{F}_6^{\mathcal{T}|j_1\cdots j_6}(\theta_1,\cdots,\theta_6)
\left({F}_4^{\mathcal{T}|j'_1\cdots j'_4}(\theta'_1,\cdots,\theta'_4)\right)^*\\\nonumber
&\times e^{-\mathrm{m} r_L(\cosh\theta_1+\cdots\cosh\theta_6)-\mathrm{m} r_R(\cosh\theta'_1+\cdots\cosh\theta'_4)}
\end{align}
After integrating out all the delta functions and some rearrangement, we obtain
\begin{align}
\label{eq:example2p}
\texttt{term}=&\,\frac{1}{6!\,4!}\sum_{j_1,j_2=1}^n\sum_{j_5,j_6=1}^n\sum_{j'_3=1}^n\int\frac{d\theta_1 d\theta_2 d\theta_5 d\theta_6 d\theta'_3}{(2\pi)^5}
\,\hat{R}(\theta_5)\hat{R}(\theta_6)\hat{T}(\theta_1)\hat{T}(\theta_2)\hat{R}(\theta'_3)\,(-1)^{\delta_{j_5,j_6}}\\\nonumber
&\,\times {F}_6^{\mathcal{T}|j_1\,j_2\,j_5\,j_6\,j_5\,j_6}(\theta_1,\theta_2,-\theta_5,-\theta_6,\theta_5,\theta_6)
\left({F}_4^{\mathcal{T}|j_1\,j_2\,j'_3\,j'_3}(\theta_1,\theta_2,\theta'_3,-\theta'_3)\right)^*\\\nonumber
&\,\times e^{-2\mathrm{m}r_L(\cosh\theta_5+\cosh\theta_6)-2\mathrm{m}r_R\cosh\theta'_3-\mathrm{m}r(\cosh\theta_1+\cosh\theta_2)}
\end{align}
In what follows, as a convention, we denote the particles that are connected by blue lines by $\{\mu_a,k_a\}$. Note that in (\ref{eq:example2p}) we can rewrite $F_6^{\mathcal{T}}$ as
\begin{align}
{F}_6^{\mathcal{T}|j_1\,j_2\,j_5\,j_6\,j_5\,j_6}(\theta_1,\theta_2,-\theta_5,-\theta_6,\theta_5,\theta_6)
=(-1)^{\delta_{j_5,j_6}}{F}_6^{\mathcal{T}|j_1\,j_2\,j_5\,j_5\,j_6\,j_6}(\theta_1,\theta_2,-\theta_5,\theta_5,-\theta_6,\theta_6)
\end{align}
where the factor $(-1)^{\delta_{j_5,j_6}}$ cancels exactly the same factor in (\ref{eq:example2p}). Relabeling $\{\theta_5,j_5\}$ and $\{\theta_6,j_6\}$ and $\{\theta'_3,j'_3\}$, we can write
\begin{align}
\label{eq:example2}
\texttt{term}=&\,\frac{1}{6!\,4!}\sum_{j_1,j_2=1}^n\sum_{k_1,k_2=1}^n\sum_{j'_1=1}^n\int\frac{d\theta_1d\theta_2 d\mu_1 d\mu_2 d\theta'_1}{(2\pi)^5}
\,\hat{R}(\theta_1)\hat{R}(\theta_2)\hat{T}(\mu_1)\hat{T}(\mu_2)\hat{R}(\theta'_1)\\\nonumber
&\,\times {F}_6^{\mathcal{T}|j_1\,j_1\,j_2\,j_2\,k_1\,k_2}(-\theta_1,\theta_1,-\theta_2,\theta_2,\mu_1,\mu_2)
{F}_4^{\mathcal{T}|j'_1\,j'_1\,k_1\,k_2}(-\theta'_1,\theta'_1,-\mu_1,-\mu_2)\\\nonumber
&\,\times e^{-2\mathrm{m} r_L(\cosh\theta_1+\cosh\theta_2)-2\mathrm{m}r_R\cosh\theta'_1-\mathrm{m}r(\cosh\mu_1+\cosh\mu_2)}
\end{align}
where we have used (\ref{eq:Fconj}). Let us also notice that using Waston's equation, we can move any pairs in the form factor
\begin{align}
F_N^{\mathcal{T}|\cdots j\,j_a\,j_a\cdots}(\cdots,\textcolor{blue}{\theta},-\theta_a,\theta_a,\cdots)=
F_N^{\mathcal{T}|\cdots j_a\,j_a\,j\cdots}(\cdots,-\theta_a,\theta_a,\textcolor{blue}{\theta},\cdots)
\end{align}
Therefore after integrating out the delta functions, we can always bring the form factors into the form
\begin{align}
F_{2N+M}^{\mathcal{T}|j_1j_1\cdots j_Nj_N k_1\cdots k_M}(-\theta_1,\theta_1,\cdots,-\theta_N,\theta_N,\mu_1,\cdots,\mu_M).
\end{align}
Using the procedure described so far, one can show that all the possible diagrams with $6|4$ particles on the $\text{left}|\text{right}$ and two transmissive particles give the same result as the diagram shown in figure\,\ref{fig:dipole}.
We call this kind of diagram the `standard diagram' and denote the standard diagram with $M$ particles on the left and $N$ particles on the right of the defect with $k$ pairs of transmissive particles as $\rT_{M,N}^{(k)}$. For each term $f_{M,N}$ in the spectral expansion, we only need to consider the standard diagram $\rT_{M,N}^{(k)}$ with $k=0,\cdots,\text{min}(M,N)$. The expression for each standard diagram can be easily written down by the rules given above. It is obvious that the only non-vanishing standard diagrams are $\rT_{2m,2m'}^{(2a)}$, the multiplicity of which can be computed as
\begin{align}
\label{eq:cmma}
c_{m,m'}^a=\frac{1}{2^{m+m'-2a}}\frac{(2m)!(2m')!}{(m-a)!(m'-a)!(2a)!}.
\end{align}
To conclude, the computation of each term $f_{2m,2m'}$ in the spectral expansion is given by
\begin{align}
f_{2m,2m'}=\frac{1}{(2m)!(2m')!}\sum_{a=0}^{\text{min}[m,m']}c_{m,m'}^{a}\,\rT_{2m,2m'}^{(2a)}
\end{align}
where $\rT_{2m,2m'}^{(2a)}$ is the standard diagram whose explicit expression is given by
\begin{align}
\label{eq:Tmma}
\rT_{2m,2m'}^{(2a)}=&\,\frac{1}{(2m)!(2m')!}\sum_{j_1,\cdots,j_{m-a}}\sum_{j'_1,\cdots,j'_{m'-a}}\sum_{k_1,\cdots,k_{2a}}
\int_{-\infty}^\infty\prod_{r=1}^{m-a}\frac{d\theta_r}{2\pi}\prod_{s=1}^{m'-a}\frac{d\theta'_s}{2\pi}\prod_{t=1}^{2a}\frac{d\mu_t}{2\pi}\\\nonumber
&\,\times F_{2m}^{\mathcal{T}|j_1j_1\cdots j_{m-a}j_{m-a}k_1\cdots k_{2a}}(-\theta_1,\theta_1,\cdots,-\theta_{m-a},\theta_{m-a},\mu_1,\cdots,\mu_{2a})\\\nonumber
&\,\times F_{2m'}^{\mathcal{T}|j'_1j'_1\cdots j'_{m'-a}j'_{m'-a}k_1\cdots k_{2a}}(-\theta'_1,\theta'_1,\cdots,-\theta'_{m'-a},\theta'_{m'-a},-\mu_1,\cdots,-\mu_{2a})\\\nonumber
&\,\times \prod_{r=1}^{m-a}\hat{R}(\theta_r)e^{-2\mathrm{m} r_L\cosh\theta_r}\prod_{s=1}^{m'-a}\hat{R}(\theta'_s)e^{-2\mathrm{m} r_R\cosh\theta'_s}
\prod_{t=1}^{2a}\hat{T}(\mu_t)e^{-\mathrm{m} r\cosh\mu_t}
\end{align}
and $c_{m,m'}^a$ is given in (\ref{eq:cmma}). As a consistency check, we notice that if we restrict to the sums of $\rT_{2m,0}^{(0)}$ or $\rT_{0,2m}^{(0)}$, the results are exactly the spectral expansion of boundary EE studied in \cite{CastroAlvaredo:2008pf}. If we restrict to the sum of $\rT_{2m,2m}^{(2m)}$, we obtain the spectral expansion of EE in the presence of a toloplogical defect studied in \cite{Jiang:2017qhn}. Further putting $\hat{T}(\theta)=1$ gives the spectral expansion of the bulk EE.\par

To summarize, after plugging the defect matrix elements into the spectral expansion and integrating out the delta functions, we can write the defect two-point function as
\begin{align}
\label{eq:spectralexp}
\langle\mathcal{T}(t_1)\mathbb{D}(t_\rD)\tilde{\mathcal{T}}(t_2)\rangle
=&\,\sum_{m,m'=1}^\infty\sum_{a=0}^{\min[m,m']}\frac{c_{m,m'}^a}{(2m)!(2m')!}\rT_{2m,2m'}^{(2a)}-1\\\nonumber
&\,+\sum_{m=0}^\infty\frac{c_m}{(2m)!}\rT_{2m,0}+\sum_{m'=0}^\infty\frac{c_{m'}}{(2m')!}\rT_{0,2m'}
\end{align}
where $c_m=(2m)!/(2^m m!)$ and
\begin{align}
\frac{c_m}{(2m)!}\rT_{2m,0}=&\,\frac{1}{(4\pi)^m m!}\sum_{j_1,\cdots,j_{m}=1}^n
\prod_{r=1}^{m}\left[\int_{-\infty}^\infty \hat{R}(\theta_r)e^{-2\mathrm{m} r_L\cosh\theta_r}d\theta_r\right]\\\nonumber
&\times F_{2m}^{\mathcal{T}|j_1j_1\cdots j_{m}j_{m}}(-\theta_1,\theta_1,\cdots,-\theta_{m},\theta_{m})\\\nonumber
\frac{c_m'}{(2m')!}\rT_{0,2m'}=&\,\frac{1}{(4\pi)^{m'} m'!}\sum_{j_1,\cdots,j_{m'}=1}^n
\prod_{r=1}^{m'}\left[\int_{-\infty}^\infty \hat{R}(\theta_r)e^{-2\mathrm{m} r_R\cosh\theta_r}d\theta_r\right]\\\nonumber
&\times F_{2m'}^{\mathcal{T}|j_1j_1\cdots j_{m'}j_{m'}}(-\theta_1,\theta_1,\cdots,-\theta_{m'},\theta_{m'})
\end{align}
The last two terms on the r.h.s of (\ref{eq:spectralexp}) take exactly the same form as the spectral expansions of $\langle0|\mathcal{T}|\mathcal{B}\rangle$ and $\langle\mathcal{B}|\tilde{\mathcal{T}}|0\rangle$ \cite{CastroAlvaredo:2008pf} where $|\mathcal{B}\rangle$ is the integrable boundary state. In the purely reflective case, the first term in (\ref{eq:spectralexp}) factorizes into a product due to the fact $c_{m,m'}^{a=0}=c_m\,c_{m'}$ and $\rT_{2m,2m'}=\rT_{2m,0}\,\rT_{0,2m'}$. As a result, (\ref{eq:spectralexp}) becomes $\langle0|\mathcal{T}|\mathcal{B}\rangle\langle\mathcal{B}|\tilde{\mathcal{T}}|0\rangle$ which is nothing but the product of two boundary one-point functions.

\subsection{Fully connected terms}
In this subsection, we elaborate on the computation of $\rT_{2m,2m'}^{(2a)}$ and show that only some special terms called \emph{fully connected terms} in \cite{CastroAlvaredo:2008pf} contribute to the entanglement entropy. We then classify all inequivalent fully connected terms and compute their corresponding multiplicities.\par

\begin{figure}[h!]
\begin{center}
\includegraphics[scale=0.3]{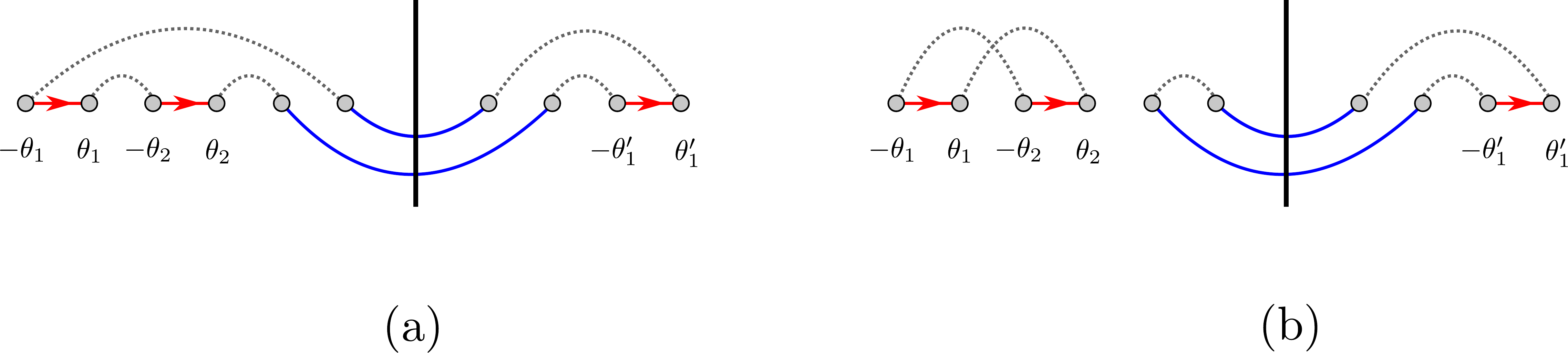}
\caption{Diagrammatic representation of terms of $F_6^{\mathcal{T}}\times F_4^{\mathcal{T}}$. Each dashed line represent a term $K_{ij}$. When two dashed lines cross, we multiply by an extra $(-1)$. Diagram (a) corresponds to a fully connected term while diagram (b) corresponds to a term which factorizes into two parts.}
\label{fig:RT1}
\end{center}
\end{figure}

For computing the standard diagram $\rT_{2m,2m'}^{(2a)}$, the most non-trivial piece is the product of two form factors together with the summation over all replica indices
\begin{align}
&\,\sum_{j_1,\cdots,j_{m-a}=1}^n\sum_{j'_1,\cdots,j'_{m'-a}=1}^n\sum_{k_1,\cdots,k_{2a}=1}^n\\\nonumber
&\,\phantom{\times}F_{2m}^{\mathcal{T}|j_1j_1\cdots j_{m-a}j_{m-a}k_1\cdots k_{2a}}(-\theta_1,\theta_1,\cdots,-\theta_{m-a},\theta_{m-a},\mu_1,\cdots,\mu_{2a})\\\nonumber
&\,\times F_{2m'}^{\mathcal{T}|j'_1j'_1\cdots j'_{m'-a}j'_{m'-a}k_1\cdots k_{2a}}(-\theta'_1,\theta'_1,\cdots,-\theta'_{m'-a},\theta'_{m'-a},-\mu_1,\cdots,-\mu_{2a})\\\nonumber
\end{align}
Let us first consider the cases $a>0$, namely the terms with non-zero number of transmissive particles. Using (\ref{eq:replicaN}) we can perform the summation over $k_{1}$ and obtain\footnote{We relabel the indices $k_i$ by $k_{i-1}$ after the summation over $k_1$.}
\begin{align}
&\,n\sum_{j_1,\cdots,j_{m-a}=0}^{n-1}\sum_{j'_1,\cdots,j'_{m'-a}=0}^{n-1}\sum_{k_1,\cdots,k_{2a-1}=0}^{n-1}\\\nonumber
&\,\phantom{\times}F_{2m}^{\mathcal{T}|11\cdots1}((-\theta_1)^{[j_1]},\theta_1^{[j_1]},\cdots,\mu_1,\mu_2^{[k_1]},\cdots,\mu_{2a}^{[k_{2a-1}]})\\\nonumber
&\,\times F_{2m'}^{\mathcal{T}|11\cdots1}((-\theta'_1)^{[j'_1]},(\theta'_1)^{[j'_1]},\cdots,-\mu_1,(-\mu_2)^{[k_1]},\cdots,(-\mu_{2a-1})^{[k_{2a-1}]})
\end{align}
where we have introduced the notation
\begin{align}
\theta_a^{[j]}=\theta_a+2\pi i j
\end{align}
and similar for $\theta'_a$ and $\mu_a$. Then we can use formula (\ref{eq:Pfaffian}) to evaluate the form factors $F_{2m}^{\mathcal{T}|1\cdots1}$ in terms of Pfaffians. The Pfaffian can be written as the sum over all possible Wick contractions of the particles of the form factors. We use dashed lines that connect two nodes to denote the Wick contractions. The Wick contraction of particles with rapidity $\theta_i$ and $\theta_j$ corresponds to a factor $K(\theta_i-\theta_j)$. We can use diagrams as in figure\,\ref{fig:RT1} to represent each term of the Pfaffian.
Among these terms, one special type is the fully connected terms as in figure\,\ref{fig:RT1} (a). They have the property that the sum over replica numbers does not factorize into two or more sums. If all the nodes of a diagram are connected (either by solid lines or dashed lines), it corresponds to a fully connected term. Figure\,\ref{fig:RT1} (b) corresponds to the product of two fully connected terms. Due to the property (\ref{eq:propertyK}), any fully connected term vanishes in the limit $n\to 1$. However, it has been shown in \cite{CastroAlvaredo:2008pf} that the derivative of a fully connected term with respect to $n$ does not vanish for $n\to 1$. Therefore, the derivatives of products of two or more fully connected terms always contain a vanishing piece in the limit $n\to1$ and thus do not contribute to the entanglement entropy. For the case $a=0$, the sum always factorizes and it is not fully connected. We thus reduce the computation of standard diagrams to the classification and computation of fully connected terms.\par

In order to characterize the fully connected terms, we first analyze a simpler quantity without the shifts of rapidities
\begin{align}
\label{eq:FFex}
&\,F_{2m}^{\mathcal{T}|1\cdots1}(-\theta_1,\theta_1,\cdots,-\theta_{m-a},\theta_{m-a},\mu_1,\cdots,\mu_{2a})\\\nonumber
\times &\, F_{2m'}^{\mathcal{T}|1\cdots1}(-\theta'_1,\theta'_1,\cdots,-\theta'_{m'-a},\theta'_{m'-a},-\mu_1,\cdots,-\mu_{2a}).
\end{align}
All the fully connected terms can be written in the following form
\begin{align}
-\rK(\mu_1,\cdots,\mu_{\sigma_1},-\mu_{\sigma_1},\cdots,-\mu_{\sigma_2},\mu_{\sigma_2},\cdots,\mu_{\sigma_{2a-1}},-\mu_{\sigma_{2a-1}},\cdots,-\mu_1)
\end{align}
where $\sigma$ is any permutation of $\{2,3,\cdots,2a\}$ and we have defined
\begin{align}
\rK(\theta_1,\theta_2,\cdots,\theta_{2N})\equiv K(\theta_1-\theta_2)K(\theta_3-\theta_4)\cdots K(\theta_{2N-1}-\theta_{2N}).
\end{align}
Between any $\mu_a,\cdots,\mu_b$, one can insert any number of the pairs $\{-\theta_j,\theta_j\}$ ($j=1,\cdots,m-a$) which we sometimes refer to as `dipoles' while between any $-\mu_a,\cdots,-\mu_b$ one can insert any number of pairs $\{-\theta'_j,\theta'_j\}$ ($j=1,\cdots,m'-a$). For example, figure\,\ref{fig:RT1} (a) can be written as
\begin{align}
-\rK(\mu_1,\textcolor{red}{-\theta_2,\theta_2,-\theta_1,\theta_1},\mu_2,-\mu_2,\textcolor{red}{-\theta'_1,\theta'_1},-\mu_1)
\end{align}
In fact, all fully connected terms of (\ref{eq:FFex}) can be constructed in three steps
\begin{enumerate}
\item Write down $\rK(\mu_1,\mu_{\sigma_1},-\mu_{\sigma_1},-\mu_{\sigma_2},\mu_{\sigma_2},\cdots,-\mu_{\sigma_{2a-1}},-\mu_1)$ without any $\theta_j,\theta'_{j'}$ where we have fixed the first and last variable to be $\mu_1$ and $-\mu_1$. We still have $(2a-1)!$ possible permutations. By relabelling the indices, all of them are equivalent.
\item Distribute all the $m-a$ pairs of $\{-\theta_1,\theta_1\},\cdots,\{-\theta_{m-a},\theta_{m-a}\}$ between the `chambers' $\mu_1,\mu_{\sigma_1}$ and $\mu_{\sigma_i},\mu_{\sigma_{i+1}}$;
\item Distribute all the $m'-a$ pairs of $\{-\theta'_1,\theta'_1\},\cdots,\{-\theta'_{m-a},\theta'_{m-a}\}$ between the `chambers' $-\mu_{\sigma_i},-\mu_{\sigma_{i+1}}$ and $-\mu_{2a-1},-\mu_1$.
\end{enumerate}
In the boundary case, all the fully connected terms are equivalent. However, this is not so for the defect case. We thus need to classify the fully connected terms in this case and compute the corresponding multiplicity. Based on the three steps described above, the fully connected terms can be characterized by two sets of numbers $(\vec{m}|\vec{m}')$ where $\vec{m}=(m_1,m_2,\cdots,m_a)$ and $\vec{m}'=(m'_1,m'_2,\cdots,m'_a)$. Each $m_i$ denotes the number of dipoles in the chamber $i$ and similar for $m'_i$. The elements of $\vec{m}$ and $\vec{m}'$ satisfy the following constraints
\begin{align}
\sum_{i=1}^a m_i=N_L,\qquad \sum_{i=1}^a m'_i=N_R,\qquad m_i,m'_i\ge0
\end{align}
where $N_L=m-a$ and $N_R=m'-a$ are the total numbers of dipoles on the left and right, respectively. Now we count the multiplicity for each configuration $(\vec{m}|\vec{m}')$. We have the freedom to permute $\mu_2,\cdots,\mu_{2a}$\footnote{As a convention, we always put $\mu_1$ at the first position.} which gives $(2a-1)!$. We also have the freedom to relabel the dipoles on the left and right, this gives $N_L!/(m_1!\cdots m_a!)$ and $N_R!/(m'_1!\cdots m'_a!)$, respectively. Finally we can flip the sign within each dipole, which gives $2^{N_L}$ and $2^{N_R}$. For each $(\vec{m}|\vec{m}')$, we only need to consider one `representative' of the following form
\begin{align}
\Gamma^a_{\vec{m}|\vec{m}'}=-\rK(\mu_1,\bm{\Theta}_1,-\mu_2,\mu_2,\bm{\Theta}'_1,-\mu_3,\mu_3,\cdots,\bm{\Theta}_a,-\mu_{2a},\mu_{2a},\bm{\Theta}'_{a},-\mu_1)
\end{align}
where we have used the fact that we can change the sign of $\mu_k$ in the integral since $\hat{T}(-\mu)=\hat{T}(\mu)$. The $\bm{\Theta}_k$ $(k=1,\cdots,a)$ are given by
\begin{align}
\bm{\Theta}_1=&\,\{-\theta_1,\theta_1,\cdots,-\theta_{m_1},\theta_{m_1}\},\\\nonumber
\bm{\Theta}_2=&\,\{-\theta_{m_1+1},\theta_{m_1+1},\cdots,-\theta_{m_1+m_2},\theta_{m_1+m_2}\},\\\nonumber
\cdots\\\nonumber
\bm{\Theta}_a=&\,\{-\theta_{m_1+\cdots+m_{a-1}+1},\theta_{m_1+\cdots+m_{a-1}+1},\cdots,-\theta_{N_L},\theta_{N_L}\}
\end{align}
so that
\begin{align}
(\bm{\Theta}_1,\cdots,\bm{\Theta}_a)=(-\theta_1,\theta_1,-\theta_2,\theta_2,\cdots,-\theta_{N_L},\theta_{N_L}).
\end{align}
If some $m_k=0$, then $\bm{\Theta}_k$ is given by an empty set. The definitions for $\bm{\Theta}'_k$ $(k=1,\cdots,a)$ are similar. For example,
\begin{align}
\Gamma^2_{(1,1|2,0)}=
-\rK(\mu_1,\textcolor{blue}{-\theta_1,\theta_1},
-\mu_2,\mu_2,\textcolor{red}{-\theta'_1,\theta'_1,-\theta'_2,\theta'_2},-\mu_3,\mu_3,\textcolor{blue}{-\theta_2,\theta_2},-\mu_4,\mu_4,
\textcolor{gray}{\emptyset},-\mu_1)
\end{align}
We can see from the example that if we denote
\begin{align}
(\lambda_1,\lambda_2,\cdots,\lambda_8)\equiv
(\mu_1,\theta_1,\mu_2,,\theta'_1,\theta'_2,\mu_3,\theta_2,\mu_4,)
\end{align}
then we can write
\begin{align}
\Gamma^2_{(1,1|2,0)}=&\,-\rK(\lambda_1,-\lambda_2,\lambda_2,\cdots,-\lambda_1)
=K(-\hat{\lambda}_{12})K(\hat{\lambda}_{23})\cdots K(\hat{\lambda}_{18})
\end{align}
After putting back the shifts from replica indices, the above quantity takes exactly the same form as the one that has been studied in \cite{CastroAlvaredo:2008pf}. This is true for all $\Gamma^a_{\vec{m}|\vec{m}'}$ and we can take advantage of this fact and apply the same analytical continuation in $n$ when we put back the shifts of rapidities
\begin{align}
\tilde{\Gamma}_{\vec{m}|\vec{m}'}^a=
-\rK(\mu_1,\tilde{\bm{\Theta}}_1,-\tilde{\mu}_2,\tilde{\mu}_2,\tilde{\bm{\Theta}}'_1,-\tilde{\mu}_3,
\tilde{\mu}_3,\cdots,\tilde{\bm{\Theta}}_a,-\tilde{\mu}_{2a},\tilde{\mu}_{2a},\tilde{\bm{\Theta}}'_{a},-\mu_1)
\end{align}
where we define
\begin{align}
\tilde{\theta}_a=&\,\theta_a+2\pi i j_a,& \widetilde{-\theta_a}=&\,-\theta_a+2\pi i j_a&\\\nonumber
\tilde{\theta'}_a=&\,\theta'_a+2\pi i j'_a,& \widetilde{-\theta'_a}=&\,-\theta'_a+2\pi i j'_a&\\\nonumber
\tilde{\mu}_a=&\,\mu_a+2\pi i k_{a-1},& \widetilde{-\mu_a}=&\,-\mu_a+2\pi i k_{a-1}.&
\end{align}
This allows us to compute the following central quantity
\begin{align}
\Upsilon_{\vec{m}|\vec{m}'}^a\equiv-\frac{d}{dn}\left[\sum_{j_1,\cdots,j_{m-a}=0}^{n-1}\,\,\sum_{j'_1,\cdots,j'_{m'-a}=0}^{n-1}\,\,
\sum_{k_1,\cdots,k_{2a-1}=0}^{n-1}\tilde{\Gamma}_{\vec{m}|\vec{m}'}^a\right]_{n=1}.
\end{align}
For each $\Upsilon_{\vec{m}|\vec{m}'}^a$, let us re-name the rapidities
\begin{align}
\{\lambda_1,\lambda_2,\cdots,\lambda_{m+m'}\}=\{\mu_1,\bm{\theta}_1,\mu_2,\bm{\theta}'_1,\cdots,\mu_{2a-1},\bm{\theta}_a,\mu_{2a},\bm{\theta}'_a\}
\end{align}
where
\begin{align}
\{\bm{\theta}_1,\bm{\theta}_2,\cdots,\bm{\theta}_a\}\equiv&\,\{\theta_1,\theta_2,\cdots,\theta_{m-a}\}\\\nonumber
\{\bm{\theta}'_1,\bm{\theta}'_2,\cdots,\bm{\theta}'_a\}\equiv&\,\{\theta'_1,\theta'_2,\cdots,\theta'_{m'-a}\}
\end{align}
and the order of particles in the sets should be respected. Using the prescription for analytic continuation proposed in \cite{CastroAlvaredo:2008pf}, when $m+m'=2\ell$, we have
\begin{align}
\label{eq:Upeven}
\Upsilon_{\vec{m}|\vec{m}'}^a\mapsto &\, \hat{\Upsilon}_{\vec{m}|\vec{m}'}^a= -2\pi^2\sum_{j=1}^{\ell}\sum_{k=1}^j\sum_{\epsilon=\pm}(-1)^{\ell+j}a_{k,j}\mathrm{C}_{2\ell-1}^{\ell-j}\,\\\nonumber
&\,\times\dfrac{\delta(\Lambda)\displaystyle\prod_{r=1}^{2\ell}\exp{\left(\frac{\epsilon(j-k)\pi i}{2\ell}\partial_{\lambda_r}\right)}}
{\displaystyle\prod_{r=1}^{2\ell}2\cosh\left(\frac{\hat{\lambda}_{r,r+1}}{2}+\frac{\epsilon(j-k)\pi i}{2\ell}  \right)}
\end{align}
%\text{shift}_{\lambda_r\to\lambda_r+\frac{\epsilon(j-k)\pi i}{2\ell}}
where $a_{j,k}=1$ for $k<j$ and $a_{j,j}=1/2$ and $\mathrm{C}_m^n$ denotes the binomial number $\mathrm{C}_m^n=m!/(n!(m-n)!)$. The argument of the $\delta$-function is given by the sum of all rapidities
\begin{align}
\Lambda=\sum_{i=1}^{2\ell}\lambda_i
\end{align}
The arrow in (\ref{eq:Upeven}) means we replace $\Upsilon_{\vec{m}|\vec{m}'}^a$ by the quantity on the r.h.s. inside the integral and the shift operator $\exp(x\,\partial_{\lambda_r})$ shifts the argument $\lambda_r$ of the remaining function (this includes the transmission and reflection amplitudes and the exponential factors) by amount of $x$. When $m+m'=2\ell+1$, we have
\begin{align}
\label{eq:Upodd}
\Upsilon_{\vec{m}|\vec{m}'}^a\mapsto\,\hat{\Upsilon}_{\vec{m}|\vec{m}'}^a=
(-1)^{\ell}\,2\pi^2\,\delta(\Lambda)\sum_{j=1}^{\ell}\sum_{k=1}^j\sum_{\epsilon=\pm}
\frac{\mathrm{C}_{2\ell}^{\ell-j}(-1)^j\,\epsilon\displaystyle\prod_{r=1}^{2\ell+1}\exp{\left(\frac{\epsilon(j-k+1/2)i\pi}{2\ell+1}\partial_{\lambda_r}\right)}}
{\displaystyle\prod_{r=1}^{2\ell+1}2\cosh\left(\frac{\hat{\lambda}_{r,r+1}}{2}+\frac{\epsilon(j-k+1/2)\pi i}{2\ell+1}  \right)}
\end{align}
The equations (\ref{eq:Upeven}) and (\ref{eq:Upodd}) give the results of fully connected terms that are associated with the configuration $(\vec{m}|\vec{m}')$. The multiplicity of the fully connected terms of the type $(\vec{m}|\vec{m}')$ with $a>0$ is
\begin{align}
d^a_{\vec{m}|\vec{m}'}=2^{m+m'-2a}(2a-1)!(m-a)!(m'-a)!
\end{align}
which is independent of $\vec{m}$ and $\vec{m}'$. The multiplicity for the purely reflective fully connected terms labeled by $(m|0)$ and $(0|m)$ are given by \cite{CastroAlvaredo:2008pf}
\begin{align}
d_m=2^{m-1}(m-1)!
\end{align}

\subsection{Final result}
In this section, we combine the pieces that we obtained so far and write down the final result. In the IR limit,
\begin{align}
\label{eq:Sexpand}
S_A^{\text{defect}}=-\frac{c}{3}\log(\mathrm{m}r)+U+\sum_{m,m'=1}^\infty s_{2m|2m'}+\sum_{m=0}^\infty s_{2m|0}+\sum_{m'=0}^\infty s_{0|2m'}
\end{align}
where the three infinite sums in (\ref{eq:Sexpand}) come from the corresponding sums in (\ref{eq:spectralexp}) and the constant $U$ is the same one defined in the bulk case. The explicit form of $s_{2m|2m'}$ is given by
\begin{align}
s_{2m|2m'}=&\,\sum_{a=1}^{\text{min}[m,m']}\frac{c_{m,m'}^a}{(2m)!(2m')!}\int_{-\infty}^{\infty}
\prod_{r=1}^{m-a}\frac{d\theta_r}{2\pi}\prod_{s=1}^{m'-a}\frac{d\theta'_s}{2\pi}\prod_{t=1}^{2a}\frac{d\mu_t}{2\pi}\,
\\\nonumber
&\,\times\sum'_{\vec{m},\vec{m}'}d^a_{\vec{m}|\vec{m}'}\,\hat{\Upsilon}_{\vec{m}|\vec{m}'}^a\left[ \prod_{r=1}^{m-a}\hat{R}(\theta_r)e^{-2\mathrm{m}r_L\cosh\theta_r}\prod_{s=1}^{m'-a}\hat{R}(\theta'_s)e^{-2\mathrm{m}r_R\cosh\theta'_s}
\prod_{t=1}^{2a}\hat{T}(\mu_t)e^{-\mathrm{m}r\cosh\mu_t}\right]
\end{align}
where the prime on the summation over $\vec{m}$ and $\vec{m}'$ denotes the fact that the summation is constraint by the following relations
\begin{align}
\sum_{k=1}^a m_k=m-a,\qquad \sum_{k=1}^am'_k=m'-a.
\end{align}
We use $\hat{\Upsilon}^a_{\vec{m}|\vec{m}'}$ to denote the r.h.s. of (\ref{eq:Upeven}) and (\ref{eq:Upodd}) which contains shift operators acting on the remaining functions. This result is explicit although still quite involved. For later purpose, it is more convenient to re-name the rapidities by $\lambda_j$, $j=1,\cdots,m+m'$. There are three types of $\lambda_k$'s, which corresponds to $\mu_j,\theta_j,\theta'_j$ respectively. Let us denote the set of corresponding indices as $\alpha,\beta,\gamma$ such that
\begin{align}
\{\lambda_i|i\in\alpha\}=&\,\{\mu_1,\cdots,\mu_{2a}\},\\\nonumber
\{\lambda_j|j\in\beta\}=&\,\{\theta_1,\cdots,\theta_{m-a}\},\\\nonumber
\{\lambda_k|k\in\gamma\}=&\,\{\theta'_1,\cdots,\theta'_{m'-a}\}.
\end{align}
Then the integrand can be written as
\begin{align}
\label{eq:slambda}
s_{2m|2m'}=&\,\sum_{a=1}^{\text{min}[m,m']}\frac{1}{2a}\int_{-\infty}^{\infty}
\prod_{r=1}^{m+m'}\frac{d\lambda_r}{2\pi}\,\sum'_{\vec{m},\vec{n}}\hat{\Upsilon}_{\vec{m}|\vec{n}}^a
\\\nonumber
&\,\times\left[ \prod_{r\in\alpha}\hat{T}(\lambda_r)e^{-\mathrm{m}r\cosh\lambda_r}
\prod_{s\in\beta}\hat{R}(\lambda_s)e^{-2\mathrm{m} r_L\cosh\lambda_s}
\prod_{t\in\gamma}\hat{R}(\lambda_t)e^{-2\mathrm{m} r_R\cosh\lambda_t}
\right]
\end{align}
where we have used the fact that
\begin{align}
\frac{c_{m,m'}^a\,d^a_{\vec{m}|\vec{n}}}{(2m)!(2m')!}=\frac{1}{2a}
\end{align}
Note that configuration $(\vec{m}|\vec{m}')$ fixes uniquely the set $(\alpha,\beta,\gamma)$. As an example we take $\vec{m}=\{1,1\}$ and $\vec{m}'=\{2,0\}$, then we have
\begin{align}
\alpha=\{1,3,6,8\},\qquad \beta=\{2,7\},\qquad \gamma=\{4,5\}.
\end{align}
For completeness, we also write down the expressions for the boundary like terms
\begin{align}
\label{eq:ssboundary}
s_{2m|0}=&\,\frac{1}{2m}\int\prod_{i=1}^{m}\frac{d\theta_i}{2\pi}\,\hat{\Upsilon}_{2m}\left[\prod_{i=1}^m \hat{R}(\theta_i)e^{-2\mathrm{m}r_L\cosh\theta_i} \right]\\\nonumber
s_{0|2m'}=&\,\frac{1}{2m'}\int\prod_{i=1}^{m'}\frac{d\theta_i}{2\pi}\,\hat{\Upsilon}_{2m'}\left[\prod_{i=1}^{m'} \hat{R}(\theta_i)e^{-2\mathrm{m}r_R\cosh\theta_i}  \right]
\end{align}
where the explicit form of operator $\hat{\Upsilon}_{2m}$ can be found in \cite{CastroAlvaredo:2008pf}. For the reader's convenience, we quote the expressions here. For $m=2\ell$,
\begin{align}
\hat{\Upsilon}_{2m}=(-1)^{\ell}2\pi^2\delta(\theta)\left[\frac{\rC_{2\ell-2}^{\ell-1}}{\prod_{j=1}^{2\ell}2\cosh\frac{\hat{\theta}_{j,j+1}}{2}}  -\sum_{j=1}^{\ell}\sum_{k=1}^{j-1}\sum_{\epsilon=\pm}\frac{\rC_{2\ell-1}^{\ell-j}(-1)^j\prod_{r=1}^{2\ell}\exp\left(\epsilon\frac{j-k}{2\ell}\pi i\partial_{\theta_r} \right)}{\prod_{r=1}^{2\ell}2\cosh\left(\frac{\hat{\theta}_{r,r+1}}{2}+\epsilon\frac{j-k}{2\ell}\pi i  \right)}\right]
\end{align}
For $m=2\ell+1$,
\begin{align}
\hat{\Upsilon}_{2m}=(-1)^{\ell}2\pi^2\delta(\theta)\sum_{j=1}^{\ell}\sum_{k=1}^j\sum_{\epsilon=\pm}
\frac{\rC_{2\ell}^{\ell-j}(-1)^j\epsilon\prod_{r=1}^{2\ell+1}\exp\left(\epsilon\frac{j-k+1/2}{2\ell+1}\pi i\partial_{\theta_r}\right)}{\prod_{r=1}^{2\ell}2\cosh\left(\frac{\hat{\theta}_{r,r+1}}{2}+\epsilon\frac{j-k+1/2}{2\ell+1}\pi i\right)}
\end{align}
where $\hat{\theta}_{i,i+1}=\theta_i+\theta_{i+1}$ and $\theta=\theta_1+\cdots+\theta_m$.

%%%%%%%%%%%%%%%%%%%%%%%%%%%%%%%%%%%%%%%%%%%%%%%%%%%%%%%%%%%%%%
\section{Asymptotic behaviors}
\label{sec:asymp}
%%%%%%%%%%%%%%%%%%%%%%%%%%%%%%%%%%%%%%%%%%%%%%%%%%%%%%%%%%%%%%
In this section, we consider asymptotic behaviors of the defect EE given by (\ref{eq:Sexpand}) and (\ref{eq:slambda}),(\ref{eq:ssboundary}). Due to the presence of the defect, there are different cases we can consider. We can consider the UV limit where $r_L,r_R\to 0$. This limit gives the information of the underlying CFT and the corresponding defect. We can also consider the IR limit where $r_L,r_R\to\infty$. This is the limit where the form factor approach works best and the full result can be approximated well by the first few leading terms. Finally, we can have a mixture of the two. Namely, we can take the limit $r_L\to 0$, $r_R\to\infty$ or $r_L\to \infty$ and $r_R\to 0$. We investigate all these cases in this section and give explicit results for each cases.

\subsection{UV limit}
\label{sec:asymUV}
Let us start by considering the UV limit. There are two kinds of interesting quantities that we want to identify in this limit. The first kind are the coefficients in front of the logarithmic divergent terms. In the purely bulk case, there is only one coefficient which is given by $c/3$ where $c$ is the central charge of the corresponding CFT and in the presence of topological defect, this coefficient is not modified. However, in our case, the defect is non-topological and the coefficients become non-trivial functions depending on the parameter of the defect $\chi$. In the infinite series expansion of the entanglement entropy, we have three infinite sums and correspondingly we have three logarithmic divergences $\log(r/\varepsilon)$, $\log(2r_L/\varepsilon)$ and $\log(2r_R/\varepsilon)$. We denote the corresponding coefficients in front of these divergences as $f(\chi)$, $f_L(\chi)$ and $f_R(\chi)$ respectively. Another quantity is the defect/boundary entropy or $g$-function which characterizes the degrees of freedom living on the defect/boundary. We denote the three $g$-functions which originate from the three infinite sums as $g(\chi)$, $g_L(\chi)$ and $g_R(\chi)$.\par

Let us first outline the strategy of taking the UV limit. The boundary like infinite sums have been analyzed in \cite{CastroAlvaredo:2008pf} and we mainly focus on the double sum (\ref{eq:slambda}). One crucial point is to note that we need to distinguish the cases where $m+m'$ is even and odd since they have different asymptotic behaviors. If $m+m'$ is even, the corresponding terms have logarithmic divergence and contributes to $f(\chi)$. On the other hand, if $m+m'$ is odd, the UV limit is finite and the terms contribute to the $g$-function $g(\chi)$.\par

We first consider the case $m+m'=2\ell$. Integrating out $\lambda_{2\ell}$ using the $\delta$-function $\delta(\Lambda)$ in $\hat{\Upsilon}_{\vec{m}|\vec{m}'}^a$. We are left with the variables $\lambda_1,\cdots,\lambda_{2\ell-1}$. It is more convenient to change the variables to $x_1,\cdots,x_{2\ell-2},\lambda_{2\ell-1}$ where $x_i=\lambda_i+\lambda_{i+1}$, $i=1,\cdots,2\ell-2$. Then the $\lambda_i$ can be expressed in terms of the new variables as
\begin{align}
\label{eq:changevariable}
\lambda_i=\sum_{j=i}^{2\ell-2}(-1)^{j-i}x_j+(-1)^{1+i}\lambda_{2\ell-1},\qquad i=1,\cdots,2\ell-2.
\end{align}
We first consider the damping factors in $\hat{\Upsilon}_{\vec{m}|\vec{m}'}^a$. Neglecting the shifts of the rapidities, it takes the form
\begin{align}
\prod_{r=1}^{2\ell}\frac{1}{\cosh[\frac{1}{2}(\lambda_r+\lambda_{r+1})]}
\end{align}
After integrating out $\lambda_{2\ell}$ and using the new set of variables, we obtain
\begin{align}
\prod_{r=1}^{2\ell-2}\frac{1}{\cosh\frac{x_r}{2}}\cdot\frac{1}{\cosh\left(\frac{1}{2}\sum_{j=2}^{\ell-1}x_{2j}\right)}
\frac{1}{\cosh\left(\frac{1}{2}\sum_{j=1}^{\ell-1}x_{2j-1}\right)}
\end{align}
where we have used
\begin{align}
\lambda_{2\ell}+\lambda_1=&\,-(\lambda_1+\cdots+\lambda_{2\ell-1})=-\sum_{j=2}^{\ell-1}x_{2j},\\\nonumber
\lambda_{2\ell-1}+\lambda_{2\ell}=&\,-(\lambda_1+\cdots+\lambda_{2\ell-2})=-\sum_{j=1}^{\ell-1}x_{2j-1}.
\end{align}
The main observation is that this damping factor is independent of the variable $\lambda_{2\ell-1}$. Now let us consider the remaining factors, namely the exponential factors and the reflection/transmission amplitudes. These factors depend on each single $\lambda_i$, $i=1,\cdots,2\ell$. After integrating out $\lambda_{2\ell}$ and change of variables (\ref{eq:changevariable}), every factor depends on $\lambda_{2\ell-1}$. Let us first focus on the product of the exponential factor.
\begin{align}
\label{eq:asymsum}
\mathrm{m}r\sum_{r\in\alpha}\cosh\lambda_r=&\,\mathrm{m}r\xi_\alpha\cosh(\textcolor{red}{\lambda_{2\ell-1}}+\eta_\alpha),\\\nonumber
2\mathrm{m}r_L\sum_{s\in\beta}\cosh\lambda_s=&\,2\mathrm{m}r_L\xi_\beta\cosh(\textcolor{red}{\lambda_{2\ell-1}}+\eta_\beta),\\\nonumber
2\mathrm{m}r_R\sum_{t\in\gamma}\cosh\lambda_t=&\,2\mathrm{m}r_R\xi_\gamma\cosh(\textcolor{red}{\lambda_{2\ell-1}}+\eta_\gamma)
\end{align}
where $\xi_{\alpha,\beta,\gamma}$ and $\eta_{\alpha,\beta,\gamma}$ are functions of $x_i$. Using the fact that $r_L+r_R=r$, we can write $r_L=yr$ and $r_R=(1-y)r$ where $0<y<1$. Summing the three sums in (\ref{eq:asymsum})
\begin{align}
\mathrm{m}r\sum_{r\in\alpha}\cosh\lambda_r+2\mathrm{m}r_L\sum_{s\in\beta}\cosh\lambda_s+2\mathrm{m}r_R\sum_{t\in\gamma}\cosh\lambda_t
=\mathrm{m}r\,\xi\cosh(\lambda_{2\ell-1}+\eta)
\end{align}
where
\begin{align}
\xi^2=C^2-S^2,\qquad \cosh\eta=\frac{C}{\xi},\qquad \sinh\eta=\frac{S}{\xi}
\end{align}
with
\begin{align}
C=&\,\xi_\alpha\cosh\eta_\alpha+2y\xi_\beta\cosh\eta_\beta+2(1-y)\xi_\gamma\cosh\eta_\gamma,\\\nonumber
S=&\,\xi_\alpha\sinh\eta_\alpha+2y\xi_\beta\sinh\eta_\beta+2(1-y)\xi_\gamma\sinh\eta_\gamma.
\end{align}
Now the exponential factor can be written compactly as $e^{-\mathrm{m}r\,\xi\,\cosh(\lambda_{2\ell-1}+\eta)}$. In the limit $\mathrm{m}r\to 0$, the behavior of this function is plotted in figure 4.4 of \cite{Jiang:2017qhn}. It is almost 1 in the interval $\log(\mathrm{m}r\xi)-\eta<\lambda_{2\ell-1}<-\log(\mathrm{m}r\xi)-\eta$ and drops to zero very quickly outside this interval. Therefore it develops a logarithmic divergence and the $\mathrm{m}r\to 0$ limit is governed by the large $\lambda_{2\ell-1}$ limit. As a result, we can replace the integral over $\lambda_{2\ell-1}$ by $-2\log(\mathrm{m}r)$.\par

We then consider the reflection and transmission amplitudes. From the expression of (\ref{eq:RThat}), we can extract the large rapidity behavior of the transmission and reflection amplitude
\begin{align}
\lim_{\lambda\to\pm\infty}\hat{T}(\lambda,\chi)=\cos\chi,\qquad \lim_{\lambda\to\pm\infty}\hat{R}(\lambda,\chi)=\mp i\sin\chi.
\end{align}
Therefore for $\lambda_i$ we have
\begin{align}
\lim_{\lambda_{2\ell-1}\to\pm\infty}\hat{R}(\lambda_{2j})=&\,\pm i\sin\chi,\qquad \lim_{\lambda_{2\ell-1}\to\pm\infty}\hat{R}(\lambda_{2j+1})=\mp i\sin\chi\\\nonumber
\lim_{\lambda_{2\ell-1}\to\pm\infty}\hat{T}(\lambda_j)=&\,\cos\chi.
\end{align}
For the boundary case, the numbers of $\lambda_{2j}$ and $\lambda_{2j-1}$ are always the same since there is only the reflection amplitude. This is slightly more complicated for the defect case. For $m+m'=2\ell$ particles with $2a$ transmissive ones, if both $m-a$ and $m'-a$ are even, the asymptotic limit leads to the factor $(\cos\chi)^{2a}(\sin\chi)^{2\ell-2a}$ while if both $m-a$ and $m'-a$ are odd, we have an extra minus sign and $-(\cos\chi)^{2a}(\sin\chi)^{2\ell-2a}$. Therefore, we can write down the UV limit of $s_{2m|2m'}$ for $m+m'=2\ell$ as
\begin{align}
\label{eq:UVs}
\lim_{\mathrm{m}r\to 0}s_{2m|2m'}=&\,\log(\mathrm{m}r)\,\sum_{a=1}^{\text{min}[m,m']}\frac{(-1)^{m-a}}{2a}\left(\sum'_{\vec{m},\vec{m}'}1\right)
(\cos\chi)^{2a}(\sin\chi)^{2\ell-2a}\\\nonumber
&\,\qquad\times\frac{(-1)^{\ell+1}}{4}\left[\mathrm{C}_{2\ell-2}^{\ell-1}\,J_\ell(0)^2-\sum_{j=1}^{\ell}\sum_{k=1}^{j-1}\sum_{\epsilon=\pm}
\mathrm{C}_{2\ell-1}^{\ell-j}(-1)^j\,J_\ell(\epsilon(j-k))^2\right]
\end{align}
where
\begin{align}
J_\ell(a)=\int_{-\infty}^\infty\left[\prod_{k=1}^{\ell-1}\frac{dx_k}{4\pi}\frac{1}{\cosh\left(\frac{x_k}{2}+\frac{a\pi i}{2\ell}\right)}\right]
\frac{1}{\cosh\left(\frac{1}{2}\sum_{j=1}^{\ell-1}x_j-\frac{a\pi i}{2\ell}\right)}
\end{align}
Note that the last line of (\ref{eq:UVs}) is nothing but $2\ell g_\ell$ defined in \cite{CastroAlvaredo:2008pf}. The explicit form of $g_\ell$ is given in appendix\,\ref{app:gl}. It has been checked in \cite{CastroAlvaredo:2008pf} that
\begin{align}
\sum_{\ell=1}^\infty g_\ell=\frac{1}{6}
\end{align}
We can then simply write
\begin{align}
\lim_{\mathrm{m}r\to 0}s_{2m|2m'}=\log(\mathrm{m}r)\mathscr{F}_{m,m'}(\chi)\,g_\ell
\end{align}
where
\begin{align}
\mathscr{F}_{m,m'}(\chi)=\sum_{a=1}^{\text{min}[m,m']}\frac{(-1)^{m-a}(m+m')}{2a}(\cos\chi)^{2a}(\sin\chi)^{m+m'-2a}\,
\mathrm{C}_{m-1}^{m-a}\,\mathrm{C}_{m'-1}^{m'-a}
\end{align}
is a function of $\chi$. Here we used the simple fact that
\begin{align}
\sum'_{\vec{m},\vec{n}}1=\mathrm{C}_{m-1}^{m-a}\,\mathrm{C}_{m'-1}^{m'-a}
\end{align}
Therefore, the coefficient in front of the logarithmic divergence is given by the following infinite sum
\begin{align}
f(\chi)=\sum_{\ell=1}^\infty\sum_{m=1}^{2\ell-1}\mathscr{F}_{m,2\ell-m}(\chi)\,g_{\ell}
\end{align}
Interestingly, we have the following non-trivial identity
\begin{align}
\label{eq:identity}
\sum_{m=1}^{2\ell-1}\mathscr{F}_{m,2\ell-m}(\chi)=1-(\sin\chi)^{2\ell}
\end{align}
which enables us to write
\begin{align}
f(\chi)=\sum_{\ell=1}^\infty(1-(\sin\chi)^{2\ell})g_\ell=\frac{1}{6}-\sum_{\ell=1}^{\infty}(\sin\chi)^{2\ell}\,g_{\ell}
\end{align}
This sum is convergent for $|\sin\chi|\le 1$ and $f(\chi)$ is well defined. According to the analysis in \cite{CastroAlvaredo:2008pf}, the coefficient in front of the logarithmic divergence takes the following form
\begin{align}
f_L(\chi)=f_R(\chi)=\frac{1}{2}\sum_{\ell=1}^\infty(\sin\chi)^{2\ell}g_\ell.
\end{align}
Note that the non-trivial dependence of $\chi$ comes in because the large rapidity limit of $\hat{R}(\theta,\chi)$ is $\pm i\sin\chi$ in the defect case instead of $\pm i$ in the pure boundary case. As a result, we indeed have the relation
\begin{align}
f(\chi)+f_L(\chi)+f_R(\chi)=\frac{1}{6}.
\end{align}
This relation is due to the identity (\ref{eq:identity}), which is non-trivial and can be seen as a consistency check of our result in (\ref{eq:fff}).

\subsection{IR limit}
\label{sec:IRlimit}
Now we consider the IR limit where $\mathrm{m}r\to\infty$. In this limit, the form factor approach works best since the first few terms in the spectral expansion give very good approximation to the result. We will write down the first exponential corrections up to 4 particles. We have
\begin{align}
S_{\text{IR}}^{\text{defect}}=-\frac{c}{3}\log(\mathrm{m}\varepsilon)+U+\Delta S
\end{align}
where the constant $U$ for the Ising model has been computed in \cite{Cardy:2007mb}
\begin{align}
U=-\frac{d}{dn}\left[\mathrm{m}^{-4\Delta_n}\langle\mathcal{T}\rangle^2\right]_{n=1}=-0.131984\cdots
\end{align}
and
\begin{align}
\Delta S\approx (s_{2|0}+s_{4|0})+(s_{0|2}+s_{0|4})+s_{2|2}.
\end{align}
The first two terms are boundary like and only contains reflecting particles. They have been considered in \cite{CastroAlvaredo:2008pf}. We can simply write down the result
\begin{align}
s_{2|0}=&\,\frac{i}{8}\int_{-\infty}^{\infty}d\theta\,\frac{\tanh\theta}{\cosh\theta}\,\hat{R}(\theta)\,e^{-2\mathrm{m}r_L\cosh\theta}
=\frac{1}{4}\int_0^{\infty}\frac{(\sinh\theta)^2\sin\chi}{(\cosh\theta)^2(\cosh\theta-\sin\chi)}e^{-2\mathrm{m}r_L\cosh\theta}d\theta,\\\nonumber
s_{4|0}=&\,\frac{1}{32}\int_{-\infty}^{\infty}d\theta\,\hat{R}^2(\theta)\,e^{-4\mathrm{m}r_L\cosh\theta}
=\frac{1}{16}\int_0^{\infty}\left( \frac{\sin\chi\sinh\theta}{\cosh\theta-\sin\chi}\right)^2 \,e^{-4\mathrm{m}r_L\cosh\theta}
\end{align}
$s_{0|2}$ and $s_{0|4}$ are given by the same expressions with $r_L$ replaced by $r_R$. The term $s_{2|2}$ contains only transmissive particles and gives a contribution similar to the topological defect studied in \cite{Jiang:2017qhn}
\begin{align}
s_{2|2}=-\frac{1}{16}\int_{-\infty}^{\infty}d\lambda\,\hat{T}(\lambda)^2\,e^{-2\mathrm{m}r\cosh\lambda}
=-\frac{1}{8}\int_0^\infty\left(\frac{\cos\chi\cosh\theta}{\cosh\theta-\sin\chi}\right)^2\,e^{-2\mathrm{m}r\cosh\lambda}
\end{align}
We notice that $s_{4|0}$, $s_{0|4}$ and $s_{2|2}$ are in fact divergent in the limit $\mathrm{m}r\to 0$, which is compatible with the discussions in the previous subsection.

\subsection{Mixed case}
Let us first consider the case $r_L\to 0$ and $r_R\to\infty$. In this limit, we expect the following behavior of the correlation function
\begin{align}
\lim_{r_L\to 0\atop r_R\to\infty}\langle\mathcal{T}(t_1)\mathbb{D}(t_\rD)\tilde{\mathcal{T}}(t_2)\rangle\sim\frac{\mathcal{N}_L(r_R)}{(2r_L)^{\Delta_L}}
\end{align}
The entanglement entropy in this limit behaves as
\begin{align}
\label{eq:mix}
\lim_{r_L\to 0\atop r_R\to\infty}S_A^{\text{defect}}=&\,-\lim_{n\to 1}\frac{d}{dn}\mathcal{Z}_n\left(\frac{\varepsilon}{2r_L}\right)^{\Delta_L}
\left(\varepsilon^{4\Delta_n-\Delta_L}\mathrm{m}^{4\Delta_n-\Delta_L} \right)\left(\mathrm{m}^{-4\Delta_n}\langle\mathcal{T}\rangle^2\right)
\left(\frac{\mathcal{N}_L(r_R)\mathrm{m}^{\Delta_L}}{\langle\mathcal{T}\rangle^2}\right)\\\nonumber
=&\,f_L(\chi)\log\left(2r_L/\varepsilon\right)-\left(\frac{c}{3}-f_L(\chi)\right)\log(\mathrm{m}\varepsilon)+g_L(\chi)+U+\mathcal{O}\big(e^{-a\mathrm{m} r_R},\frac{1}{(\mathrm{m} r_L)^b}\big)\\\nonumber
=&\,-\frac{c}{3}\log(\mathrm{m}\varepsilon)+f_L(\chi)\log(2\mathrm{m}r_L)+g_L(\chi)+U+\cdots
\end{align}
where we neglect the terms that decay exponentially in $\mathrm{m}r_R$ as well as the terms which decay with a power law in $\mathrm{m}r_L$. As we can see explicitly, the first and the second term exhibit the leading behavior of UV and IR behavior respectively. In the last line of (\ref{eq:mix}), we rewrite the result in another way where the coefficient in front of the logarithmic divergences is the usual constant $c/3$ and the effect of defect is a term that depends on $\chi$ and $r_L$. Exchanging `$L$' and `$R$' gives the result for the other case $r_L\to\infty$ and $r_R\to0$.

%%%%%%%%%%%%%%%%%%%%%%%%%%%%%%%%%%%%%%%%%%%%%%%%%
\section{The defect $g$-function}
\label{sec:g-function}
%%%%%%%%%%%%%%%%%%%%%%%%%%%%%%%%%%%%%%%%%%%%%%%%%
In this section, we derive an exact infinite series expression for the boundary and defect $g$-functions. The boundary $g$-functions $g_L(\chi)=g_R(\chi)$ takes a similar form as in the boundary case, the only difference being that the asymptotic value of $\hat{R}(\theta)$ in our case is no longer $\pm i$ but $\pm i\sin\chi$. We have the following expansion
\begin{align}
g_L(\chi)=g_R(\chi)=\sum_{\ell=1}^\infty c_\ell(\chi)
\end{align}
with the first term given by
\begin{align}
c_1(\chi)=\frac{i}{8}\int_{-\infty}^{\infty}\frac{\tanh\theta}{\cosh\theta}\hat{R}(\theta)
=\frac{1}{8}\int_{-\infty}^{\infty}\frac{\tanh\theta\sinh\theta\sin\chi}{\cosh\theta(\cosh\theta-\sin\chi)}
\end{align}
and
\begin{align}
c_{2\ell}(\chi)=&\,\frac{\pi^2(-1)^\ell}{2\ell}\left[\prod_{k=1}^{2\ell}\int_{-\infty}^\infty\frac{d\theta_k}{4\pi}\right]\delta(\theta)
\left[\mathrm{C}_{2\ell-2}^{\ell-1}\frac{\prod_{j=1}^{2\ell}\hat{R}(\theta_j)-(\sin\chi)^{2\ell}}{\prod_{j=1}^{2\ell}\cosh\frac{\hat{\theta}_{j,j+1}}{2}}  \right.\\\nonumber
&\,\left.-\sum_{j=1}^{\ell}\sum_{k=1}^{j-1}\sum_{\epsilon=\pm1}\mathrm{C}_{2\ell-1}^{\ell-j}(-1)^j\,
\frac{\prod_{r=1}^{2\ell}\hat{R}(\theta_r+\epsilon\frac{j-k}{2\ell}\pi i)-(\sin\chi)^{2\ell}}{\prod_{r=1}^{2\ell}\cosh\left(\frac{\hat{\theta}_{r,r+1}}{2}+\epsilon\frac{j-k}{2\ell}\pi i\right)}\right]
\end{align}
and similarly
\begin{align}
c_{2\ell+1}(\chi)=&\,\frac{\pi^2(-1)^\ell}{2\ell+1}\left[\prod_{k=1}^{2\ell+1}\int_{-\infty}^\infty\frac{d\theta_k}{2\pi}\right]\delta(\theta)\\\nonumber
&\,\times\sum_{j=1}^\ell\sum_{k=1}^j\sum_{\epsilon=\pm1}\mathrm{C}_{2\ell}^{\ell-j}(-1)^j\,\epsilon\,
\prod_{r=1}^{2\ell+1}\frac{\hat{R}(\theta_r+\epsilon\frac{j-k+1/2}{2\ell+1}\pi i)}{\cosh\left(\frac{\hat{\theta}_{r,r+1}}{2}+\epsilon\frac{j-k+1/2}{2\ell+1}\pi i\right)}.
\end{align}
Now we consider the function $g(\chi)$. Similar to the boundary $g$-function, it has two kinds of contributions. For $m+m'=2\ell$, the small $r$ limit is divergent and subtracting this divergence gives a finite quantity which contributes to the function $g(\chi)$. On the other hand, when $m+m'=2\ell+1$, the small $r$ limit is finite, as we will show in what follows and this is the second kind of contribution to $g(\chi)$.\par

For $m+m'=2\ell+1$, we have the set $\{\lambda_1,\cdots,\lambda_{2\ell+1}\}$, we can integrate out $\lambda_{2\ell+1}$ and for the rest $2\ell$ rapidities, we make the change of variable $x_i=\lambda_{i}+\lambda_{i+1}$ for $i=1,\cdots,2\ell-1$ and $\lambda_{2\ell}$. Then the product of the damping factor
\begin{align}
\prod_{r=1}^{2\ell+1}\frac{1}{\cosh[\frac{1}{2}(\lambda_r+\lambda_{r+1})]}
\end{align}
can be rewritten as
\begin{align}
\prod_{r=1}^{2\ell-1}\frac{1}{\cosh\frac{x_r}{2}}\cdot\frac{1}{\cosh\frac{1}{2}\left(\sum_{j=1}x_{2j-1}-\lambda_{2\ell}\right)}
\frac{1}{\cosh\frac{1}{2}\left(\sum_{j=1}^{\ell-1}x_{2j}+\lambda_{2\ell}\right)}
\end{align}
where we have used
\begin{align}
\lambda_{2\ell}+\lambda_{2\ell+1}=&\,-\sum_{i=1}^{2\ell-1}\lambda_i=-\sum_{j=1}^\ell x_{2j-1}+\lambda_{2\ell},\\\nonumber
\lambda_{2\ell+1}+\lambda_1=&\,-\sum_{i=2}^{2\ell}\lambda_i=-\sum_{j=1}^{\ell-1}x_{2j}+\lambda_{2\ell}.
\end{align}
Note that in contrary to the $m+m'=2\ell$ case, all the variables are involved in the damping factor. Therefore the limit $\mathrm{m}r\to 0$ is no longer divergent and can be taken safely. We can simply put $r_L=r_R=0$ in each $s_{2m|2m'}$. Therefore we can write down the expression for $g(\chi)$ as
\begin{align}
g(\chi)=\sum_{m,m'=1}^\infty c_{m|m'}(\chi)
\end{align}
where for $m+m'=2\ell$,
\begin{align}
c_{m|m'}=&\,\sum_{a=1}^{\text{min}[m,m']}\frac{1}{2a}\sum'_{\vec{m},\vec{m}'}\int_{-\infty}^{\infty}
\prod_{r=1}^{2\ell}\frac{d\lambda_r}{2\pi}\\\nonumber
&\,\quad\times\hat{\Upsilon}_{\vec{m}|\vec{m}'}^a\left[\prod_{r\in\alpha}\hat{T}(\lambda_r)\prod_{s\in\beta\cup\gamma}\hat{R}(\lambda_s)
-(-1)^{m-a}(\cos\chi)^{2a}(\sin\chi)^{2\ell-2a}  \right]
\end{align}
and similarly for $m+m'=2\ell+1$
\begin{align}
c_{m|m'}=&\,\sum_{a=1}^{\text{min}[m,m']}\frac{1}{2a}\sum'_{\vec{m},\vec{m}'}\int_{-\infty}^{\infty}
\prod_{r=1}^{2\ell}\frac{d\lambda_r}{2\pi}\,
\hat{\Upsilon}_{\vec{m}|\vec{m}'}^a\left[\prod_{r\in\alpha}\hat{T}(\lambda_r)\prod_{s\in\beta\cup\gamma}\hat{R}(\lambda_s)\right]
\end{align}

%%%%%%%%%%%%%%%%%%%%%%%%%%%%%%%%%%%%%%%%%%%%%%%%%%%%
\section{Conclusions}
\label{sec:conclusion}
%%%%%%%%%%%%%%%%%%%%%%%%%%%%%%%%%%%%%%%%%%%%%%%%%%%%
We studied the effect of integrable non-topological defect on the bipartite entanglement entropy in Ising field theory using the form factor approach. The non-topological defect modifies the UV logarithmic divergences and the IR exponential corrections. We obtain an explicit expression for the coefficients in front of the logarithmic divergences as well as the boundary/defect $g$-functions in terms of infinite series. We can also write the results in such a way that the logarithmic divergent terms are the same as in the bulk case and the effect of defect on EE is an $\mathcal{O}(1)$ term which depends on the parameters $\chi,r_L,r_R$. The theory with integrable non-topological defect can be seen as an interpolation between the bulk case and the integrable boundary case in the following sense. We can tune the parameter $\chi$ continuously. If we take $\chi=0$, the defect EE reduces to the bulk EE while if we take $\chi=\pi/2$, the defect EE becomes the sum of two independent boundary EEs. For $0<\chi<\pi/2$, we have the generic defect EE result given in this paper.\par

Although the expressions we obtain for the various quantities are explicit and can be evaluated numerically, it is desirable to have a closed form expression. This might be achievable by studying directly the underlying defect CFT. It is also interesting to do the computation on the lattice and compare with our results. As we discussed in the introduction, defect EE has been studied previously in the context of CFT and critical chains, but in a different set-up. To make contact with the previous results, we can either compute defect EE with our set-up at criticality or we can generalize our method to the previously studied set-up. In both cases, the study will shed lights on the relation between defect EE at criticality and off criticality and gives us a more complete understanding of defect EE of free fermionic systems.

%%%%%%%%%%%%%%%%%%%%%%%%%%%%%%%%%%%%%%%%%
\section*{Acknowledgements}
\label{sec:conclude}
%%%%%%%%%%%%%%%%%%%%%%%%%%%%%%%%%%%%%%%%%
I would like to thank Zoltan Bajnok, Song He for helpful comments on the manuscript. This work is partially supported by the Swiss National Science Foundation through the NCCR SwissMap.

\appendix

\section{Defect operator}
\label{sec:defectO}
In this appendix, we give the defect algebra of the replica theory for Ising model which is useful to compute the defect matrix element $\rD_{M,N}$. The Zamolodchikov-Faddeev algebra reads \cite{Smirnov:1992vz}
\begin{align}
\label{eq:ZF}
A^\mu(\theta_1)A^\nu(\theta_2)=&\,(-1)^{\delta_{\mu,\nu}}A^\nu(\theta_2)A^\mu(\theta_1)\\\nonumber
A^\dagger_\mu(\theta_1)A^\dagger_\nu(\theta_2)=&\,(-1)^{\delta_{\mu,\nu}}A^\dagger_\nu(\theta_2)A^\dagger_\mu(\theta_1)\\\nonumber
A^\mu(\theta_1)A^\dagger_\nu(\theta_2)=&\,(-1)^{\delta_{\mu,\nu}}A^\dagger_\nu(\theta_2)A^\mu(\theta_1)
+2\pi\,\delta^\mu_\nu\delta(\theta_1-\theta_2)
\end{align}
where $\mu,\nu,\rho,\sigma$ are replica numbers. The algebra between ZF operators and the Wick rotated defect operator $\mathbb{D}$ is given by
\begin{align}
\label{eq:defectA}
A^\dagger_\mu(\theta)\mathbb{D}=&\,\hat{R}(-\theta)A^\mu(-\theta)\mathbb{D}+\hat{T}(\theta)\mathbb{D}A^\dagger_\mu(\theta),\\\nonumber
A^\mu(\theta)\mathbb{D}=&\,\hat{R}(-\theta)A^\dagger_\mu(-\theta)\mathbb{D}+\hat{T}(\theta)\mathbb{D}A^\mu(\theta),\\\nonumber
\mathbb{D}A_\mu^\dagger(\theta)=&\,\hat{R}(\theta)\mathbb{D}A^\mu(-\theta)+\hat{T}(\theta)A^\dagger_\mu(\theta)\mathbb{D},\\\nonumber
\mathbb{D}A^\mu(\theta)=&\,\hat{R}(\theta)\mathbb{D}A_\mu^\dagger(-\theta)+\hat{T}(\theta)A^\mu(\theta)\mathbb{D}.
\end{align}
Note that $\mathbb{D}$ comes from fusing defects of all replica $\mathbb{D}=\rD_1\cdots\rD_n$ and the scattering between particles and the defect is diagonal. Similar to the boundary case, the defect operator $\mathbb{D}$ can also be written in terms of ZF operators
\begin{align}
\label{eq:Dd1}
\mathbb{D}=:\exp\left(\frac{1}{4\pi}\int_{-\infty}^\infty \mathscr{D}(\theta)d\theta \right):
\end{align}
where
\begin{align}
\label{eq:Dd2}
\mathscr{D}(\theta)=&\,\sum_{\mu=1}^n\left(\hat{R}(\theta)A_\mu^\dagger(-\theta)A_\mu^\dagger(\theta)+\hat{R}(-\theta)A^\mu(\theta)A^\mu(-\theta)
\right)\\\nonumber
&\,+\sum_{\mu=1}^n\left(\hat{T}(\theta)A^\dagger_\mu(\theta)A^\mu(\theta)+\hat{T}(-\theta)A_\mu^\dagger(-\theta)A^\mu(-\theta)\right)
\end{align}
From the explicit construction (\ref{eq:Dd1}) and (\ref{eq:Dd2}) we can see that $\mathbb{D}^\dagger=\mathbb{D}$. The states and dual states are given by ZF operators as
\begin{align}
|\theta_n,\cdots,\theta_1\rangle_{\mu_n,\cdots,\mu_1}=&\,A^\dagger_{\mu_n}(\theta_n)\cdots A^\dagger_{\mu_1}(\theta_1)|0\rangle,\\\nonumber
^{\mu_1,\cdots,\mu_n}\langle\theta_1,\cdots,\theta_n|=&\,\langle0|A^{\mu_1}(\theta_1)\cdots A^{\mu_n}(\theta_n).
\end{align}
and the defect matrix element
\begin{align}
\langle\theta_m,\cdots,\theta_1|\mathbb{D}|\theta'_1,\cdots,\theta'_n\rangle\equiv
\langle0|A_{\mu_m}(\theta_m)\cdots A_{\mu_1}(\theta_1)\mathbb{D}A_{\nu_1}(\theta'_1)\cdots A_{\nu_n}(\theta'_n)|0\rangle
\end{align}
can be computed using the algebra (\ref{eq:ZF}) and (\ref{eq:defectA}).

\section{Explicit expression for $g_\ell$}
\label{app:gl}
In this appendix, we give the explicit expression for $g_\ell$ from \cite{CastroAlvaredo:2008pf}
\begin{align}
g_\ell=\frac{(-1)^{\ell+1}}{8\ell}\left[\mathrm{C}_{2\ell-2}^{\ell-1}\,J_\ell(0)^2-\sum_{j=1}^{\ell}\sum_{k=1}^{j-1}\sum_{\epsilon=\pm}
\mathrm{C}_{2\ell-1}^{\ell-j}(-1)^j\,J_\ell(\epsilon(j-k))^2\right]
\end{align}
where
\begin{align}
J_\ell(a)=\int_{-\infty}^\infty\left[\prod_{k=1}^{\ell-1}\frac{dx_k}{4\pi}\frac{1}{\cosh\left(\frac{x_k}{2}+\frac{a\pi i}{2\ell}\right)}\right]
\frac{1}{\cosh\left(\frac{1}{2}\sum_{j=1}^{\ell-1}x_j-\frac{a\pi i}{2\ell}\right)}
\end{align}
The integral $J_\ell(a)$ can be evaluated explicitly and the number $g_\ell$ is given by
\begin{align}
g_\ell=\frac{1}{8\ell}\left(2^{3-2\ell}\mathrm{C}_{2\ell-3}^{\ell-2}-\frac{h_\ell}{\pi^2(\ell-1)^2}  \right)
\end{align}
where
\begin{align}
h_\ell=\left\{
         \begin{array}{ll}
           {\mathrm{C}^{\ell-1}_{2\ell-2}}{\left(\rC^{\ell/2-1}_{\ell-2}\right)^{-2}}+2\displaystyle{\sum_{p=0}^{\ell/2-1}}\rC_{2\ell-2}^{\ell-1+2p}\left(  \rC_{\ell-2}^{\ell/2-1+p}\right)^{-2}, & \hbox{$\ell$\text{ eve};} \\
           2\displaystyle{\sum_{p=0}^{(\ell-3)/2}}\rC_{2\ell-2}^{\ell+2p}\left(\rC_{\ell-2}^{\ell/2-3/2-p}\right)^{-2}, & \hbox{$\ell$\text{ odd}.}
         \end{array}
       \right.
\end{align}

\providecommand{\href}[2]{#2}\begingroup\raggedright\endgroup

%\bibliographystyle{JHEP}
%\bibliography{yunfeng}

\end{document}